\documentclass[conference]{IEEEtran}
\IEEEoverridecommandlockouts

\usepackage{cite}
\usepackage{amsmath,amssymb,amsfonts}
\usepackage{algorithmic}
\usepackage{graphicx}
\usepackage{textcomp}
\usepackage[table]{xcolor}
\usepackage{booktabs}
\usepackage{multirow}
\usepackage{url}
\usepackage[hidelinks]{hyperref}

\makeatletter
\newcommand{\linebreakand}{%
  \end{@IEEEauthorhalign}
  \hfill\mbox{}\par\vspace{0.3\baselineskip}
  \mbox{}\hfill\begin{@IEEEauthorhalign}
}
\makeatother

\newcommand{\xname}{AMMA}

\begin{document}

\title{\xname: A Multi-Chiplet Memory-Centric\\Architecture for Low-Latency\\1M Context Attention Serving}

\author{
\IEEEauthorblockN{Zhongkai Yu}
\IEEEauthorblockA{\textit{UCSD}\\
La Jolla, CA, USA\\
zhy055@ucsd.edu}
\and
\IEEEauthorblockN{Haotian Ye}
\IEEEauthorblockA{\textit{UCSD}\\
La Jolla, CA, USA\\
h5ye@ucsd.edu}
\and
\IEEEauthorblockN{Chenyang Zhou}
\IEEEauthorblockA{\textit{Columbia University}\\
New York, NY, USA\\
cz2791@columbia.edu}
\linebreakand
\IEEEauthorblockN{Ohm Rishabh Venkatachalam}
\IEEEauthorblockA{\textit{UCSD}\\
La Jolla, CA, USA\\
ovenkatachalam@ucsd.edu}
\and
\IEEEauthorblockN{Zaifeng Pan}
\IEEEauthorblockA{\textit{UCSD}\\
La Jolla, CA, USA\\
zapan@ucsd.edu}
\and
\IEEEauthorblockN{Zhengding Hu}
\IEEEauthorblockA{\textit{UCSD}\\
La Jolla, CA, USA\\
zhh068@ucsd.edu}
\linebreakand
\IEEEauthorblockN{Junsung Kim}
\IEEEauthorblockA{\textit{Yonsei University}\\
Seoul, Republic of Korea\\
junsung.kim@yonsei.ac.kr}
\and
\IEEEauthorblockN{Won Woo Ro}
\IEEEauthorblockA{\textit{Yonsei University}\\
Seoul, Republic of Korea\\
wro@yonsei.ac.kr}
\and
\IEEEauthorblockN{Po-An Tsai}
\IEEEauthorblockA{\textit{NVIDIA}\\
Santa Clara, CA, USA\\
poant@nvidia.com}
\linebreakand
\IEEEauthorblockN{Shuyi Pei}
\IEEEauthorblockA{\textit{Samsung Semiconductor}\\
San Jose, CA, USA\\
shuyi.pei@samsung.com}
\and
\IEEEauthorblockN{Yangwook Kang}
\IEEEauthorblockA{\textit{Samsung Semiconductor}\\
San Jose, CA, USA\\
yangwook.k@samsung.com}
\and
\IEEEauthorblockN{Yufei Ding}
\IEEEauthorblockA{\textit{UCSD}\\
La Jolla, CA, USA\\
yufeiding@ucsd.edu}
}

\maketitle

\begin{abstract}
All current LLM serving systems place the GPU at the center, from production-level attention-FFN disaggregation to NVIDIA's Rubin GPU-LPU heterogeneous platform. Even academic PIM/PNM proposals still treat the GPU as the central hub for cross-device communication. Yet the GPU's compute-rich architecture is fundamentally mismatched with the memory-bound nature of decode-phase attention, inflating serving latency while wasting power and die area on idle compute units. The problem is compounded as reasoning and agentic workloads push context lengths toward one million tokens, making attention latency the primary user-facing bottleneck.

To address these inefficiencies, we present AMMA, a multi-chiplet, memory-centric architecture for low-latency long-context attention. AMMA replaces GPU compute dies with HBM-PNM cubes, roughly doubling the available memory bandwidth to better serve memory-bound attention workloads. To translate this bandwidth into proportional performance gains, we introduce (i) a logic-die microarchitecture that fully exploits per-cube internal bandwidth for decode attention under a minimal power and area budget, (ii) a two-level hybrid parallelism scheme, and (iii) a reordered collective flow that reduces intra-chip die-to-die communication overhead. We further conduct a design-space exploration over per-cube compute power and intra-chip D2D link bandwidth, providing actionable guidance for hardware designers. Evaluations show that AMMA achieves 15.5$\times$ lower attention latency and 6.9$\times$ lower energy consumption compared with the NVIDIA H100.
Our code is available at {\color{blue}\url{https://github.com/zhongkaiyu/AMMA}}.
\end{abstract}

\begin{IEEEkeywords}
Near-Memory Processing, HBM, LLM, Long-Context Attention, Chiplet, Memory-Centric Architecture
\end{IEEEkeywords}

\section{Introduction}

\begin{figure*}[t]
      \centering
      \includegraphics[width=0.99\textwidth]{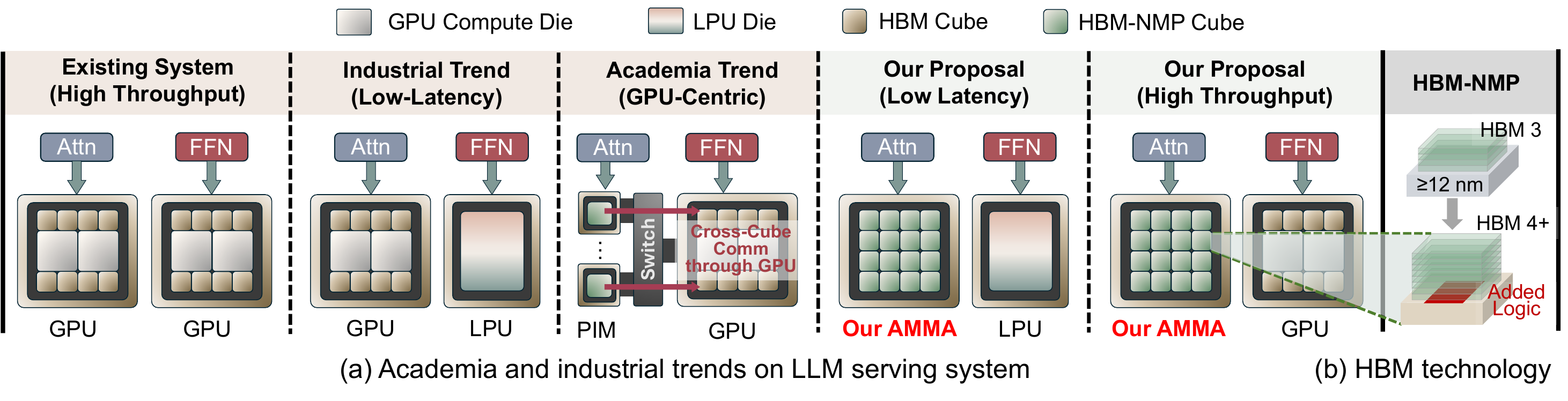}
\caption{(a) Existing serving systems rely on GPUs for decode attention. \xname{} replaces compute dies with HBM-PNM cubes to accelerate decode attention, pairing with LPUs for low-latency serving and GPUs for high-throughput serving. (b) HBM4 and beyond adopt advanced logic dies, enabling sophisticated PNM integration on the base logic die.}
      \label{fig:1_intro_disaggregation}
\end{figure*}

Existing LLM serving systems are built around a GPU-centric
paradigm.
As shown in~\autoref{fig:1_intro_disaggregation}(a),
production systems such as
MegaScale-Infer~\cite{zhu2025megascale} and
Step-3~\cite{wang2025step} adopt GPU-based
\emph{attention--FFN disaggregation}, placing attention and
FFN (MoE) layers on separate GPU pools for independent
scaling.
NVIDIA's next-generation heterogeneous platform extends this
disaggregation to the hardware level, offloading FFN layers to dedicated LPUs~\cite{groq3lpx} for ultra-low latency while retaining Rubin GPUs for attention.
Even academic proposals such as AttAcc~\cite{attacc} and
NeuPIMs~\cite{neupims}, which place processing-in-memory
(PIM) units inside DRAM to accelerate the memory-bound GEMVs
of decode attention, still rely on the GPU as the central hub
for cross-device communication.
Across industry and academia alike, the GPU remains the unchallenged centerpiece of the attention serving pipeline.

The GPU's compute-rich architecture is fundamentally mismatched
with the bandwidth-bound nature of decode attention.
This mismatch is becoming increasingly critical as reasoning and
agentic
workloads~\cite{zeng2025glm,gao2025rollpacker,pan2025kvflow,
team2025kimi,yang2024swe,yao2022webshop,fu2025areal,
wei2022chain,snell2024scaling,muennighoff2025s1} push context
lengths into the
millions~\cite{gemini15,comanici2025gemini25,claude_sonnet46,
meta2025llama4}, making decode attention latency the dominant
serving
bottleneck~\cite{zhong2024distserve,sarathi}.
To illustrate, NVIDIA Rubin provisions 795\,FLOPs per byte of
memory bandwidth~\cite{nvidia_hgx}, yet GQA attention demands
only 32\,FLOPs/byte at FP8~\cite{gqa}, a $25\times$ compute
surplus.
Rubin's two compute dies occupy 67\% of the package area and
draw 73\% of the design power, yet attention utilizes only 4\% of their
peak throughput.
With LLM serving demand already straining datacenter power
budgets~\cite{bnef2025powerai, cbre2025datacenter,
wu2026kareus}, such inefficiency is untenable.

This mismatch invites a radical rethinking of the package
architecture: \textbf{why not replace the GPU compute dies
entirely with additional HBM cubes and equip every cube with
PNM capability?}
Doing so roughly doubles the aggregated HBM bandwidth within the chip while
turning each HBM cube into a dedicated PNM accelerator.
Unlike the simple GEMV units in prior PIM/PNM
work~\cite{attacc, neupims, yun2024duplex,
li2025orches, tian2024ndpbridge, poremba2017there},
the PNM unit here is a fully functional accelerator
integrating data loading, GEMM, GEMV, and die-to-die
communication, enabling HBM cubes within the same package to
communicate directly without relying on a host GPU.
\textbf{\xname{} is therefore a standalone, fully programmable,
memory-centric accelerator with first-class system status},
representing a fundamental departure from the prevailing
GPU-centric paradigm.
It can further participate in disaggregated serving, handing
off FFN layers to LPUs for latency-sensitive deployments or
to GPUs for throughput-oriented deployments, as shown
in~\autoref{fig:1_intro_disaggregation}(a).

Fortunately, we stand at a technological inflection point that makes this vision commercially viable.
As shown in~\autoref{fig:1_intro_disaggregation}(b), starting with HBM4, memory vendors fabricate the base logic
die on advanced process nodes
(\(\le\)5\,nm)~\cite{hbm4_spec, ma2026challenges, joo202615},
providing enough transistor density for a dedicated microarchitecture that fully exploits each cube's
internal bandwidth.
Connecting such cubes via on-package D2D links then forms a
purpose-built, energy-efficient long-context attention
processor.

% \begin{figure}[t]
%       \centering
%       \includegraphics[width=0.42\textwidth]{fig/1_intro.pdf}
%             \caption{(a) Rubin GPU with compute dies and HBM cubes vs.\ our \xname{} package that replaces compute dies with HBM-NMP cubes. (b) HBM4 and later generations adopt advanced logic dies, enabling NMP added to base logic die.}
%       \label{fig:1_intro}
% \end{figure}

However, simply replacing compute dies with HBM-PNM cubes does
not automatically translate additional bandwidth into performance.
The logic-die architecture is entirely unexplored,
and the inherently distributed nature of HBM cubes makes it
challenging to fully utilize the large aggregate bandwidth
within the package. Four specific challenges arise.

\noindent\textbf{(C1) A fundamentally different compute--memory
ratio demands a new architecture.}
Each HBM cube's logic die must fully exploit its enormous
internal bandwidth under strict area and power budgets,
since the target workload is memory-bound with minimal data reuse.
This is the opposite of the GPU regime, where compute far outstrips bandwidth and a deep memory hierarchy maximizes data
reuse~\cite{hopper_white_paper, blackwell, nvidia_hgx}.
GPU architectural principles therefore do not transfer, demanding a fundamentally different design philosophy.

\noindent\textbf{(C2) Parallelism strategies introduce
long-range data movement.}
Tensor Parallelism (TP), the standard multi-GPU approach for
latency reduction, partitions attention across all
devices~\cite{shoeybi2019megatron}.
Applying TP na\"ively across 16 intra-package cubes, however,
forces communication between distant cubes separated by many
mesh hops, whose cost can easily offset the aggregated bandwidth gains.
Parallelism must instead confine communication locally.

\noindent\textbf{(C3) Multiple collective operations compound
communication overhead.}
Even with a suitable parallelism strategy, the conventional
attention flow needs multiple rounds of AllReduce and
AllGather, whose cumulative latency becomes a bottleneck,
making a simplified communication flow essential.

\noindent\textbf{(C4) Critical hardware parameters remain
unexplored.}
Two key design knobs, per-cube compute throughput and
inter-die link bandwidth, compete for shared NoC resources
under tight area and power constraints.
Determining the right balance for optimal end-to-end
attention latency requires holistic design-space exploration.

To address these challenges, we propose~\xname{},
a multi-chiplet memory-centric architecture that departs from
the current GPU-centric serving-system paradigm.
As illustrated in~\autoref{fig:1_intro_disaggregation},
\xname{} places HBM and PNM at the center,
forming a new class of LLM accelerator and opening a new direction for future, highly heterogeneous, disaggregation-based serving systems.
Concretely, \xname{} connects 16 HBM cubes via intra-package D2D
links to form a single chip.
We design a dedicated microarchitecture within each HBM's logic
die to fully exploit internal bandwidth~(\textbf{C1}).
We further introduce a two-level parallelism
scheme~(\textbf{C2}) and a reordered collective-communication
flow~(\textbf{C3}) to reduce D2D communication overhead.
Finally, we conduct a comprehensive design-space exploration to
provide actionable guidance for hardware
designers~(\textbf{C4}).
We hope this work demonstrates the potential of memory-centric
architectures as a distinct class beyond GPUs, and inspires
further research toward next-generation heterogeneous platforms
in which memory-centric accelerators play a central role.
We summarize our contributions as follows.
\begin{itemize}
    \item We propose a multi-chiplet memory-centric architecture
          for low-latency attention serving, with each cube
          integrating a dedicated microarchitecture in its logic
          die to fully utilize the aggregated HBM bandwidth.
    \item We design a two-level hybrid parallelism scheme that
          maps attention across cubes while constraining
          communication range to reduce inter-cube data movement.
    \item We redesign the collective communication flow to
          reduce both the number of collective operations and
          the global synchronization overhead, and provide a
          formal proof of correctness for the reordered flow.
    \item We explore the hardware design space by sweeping
          per-cube compute throughput and inter-die link
          bandwidth, characterizing their impact on attention
          latency and providing practical guidance for
          hardware designers.
\end{itemize}
\section{Background}

\subsection{Attention in LLM Decoding}
Each attention layer~\cite{transformer} consists of three operations.
\emph{QKV projection} linearly maps the input hidden state
$\mathbf{x} \in \mathbb{R}^{D_m}$ into queries, keys, and values via
weight matrices $\mathbf{W}_Q \in \mathbb{R}^{D_m \times H_Q d_h}$ and
$\mathbf{W}_{KV} \in \mathbb{R}^{D_m \times 2H_{KV} d_h}$,
producing $H_Q$ query heads and $H_{KV}$ key-value head pairs of
dimension $d_h$.
\emph{Core attention} is then computed per head:
\begin{equation}
  \mathbf{a}
  = \mathrm{softmax}\!\left(\frac{\mathbf{Q}\mathbf{K}^\top}{\sqrt{d_h}}\right)
    \mathbf{V},
  \label{eq:attn}
\end{equation}
where $\mathbf{Q} \in \mathbb{R}^{H_Q \times d_h}$ and
$\mathbf{K}, \mathbf{V} \in \mathbb{R}^{H_{KV} \times d_h}$.
\emph{Output projection} $\mathbf{W}_O \in \mathbb{R}^{H_Q d_h \times D_m}$
maps the concatenated head outputs back to model dimension $D_m$.
\emph{Grouped-query attention} (GQA)~\cite{gqa} reduces the KV head count
so that $G = H_Q / H_{KV}$ query heads share one KV head, lowering memory
overhead. GQA is common in modern LLMs~\cite{qwen3}.

LLM inference proceeds in two phases~\cite{sarathi,zhong2024distserve}.
\emph{Prefill} processes the entire prompt in one pass, producing
GEMM-shaped computation that is compute-bound and GPU-friendly.
\emph{Decode} generates tokens auto-regressively, one at a time.
At each decode step, the new query attends to all $S$ preceding tokens by reading their keys and values from a stored \emph{KV cache} of size $2 H_{KV} S d_h$ per layer.
With a single query token during decode, the arithmetic intensity of \autoref{eq:attn} decreases to 32~FLOPs/byte (Qwen3 $G{=}16$, FP8), far below the compute-to-bandwidth ratio of modern GPUs.
The QKV and output projections show equally low arithmetic intensity at small $B$, making the whole decode attention memory-bound.

For reasoning and agentic workloads, context length can reach the million-token regime.
{Because low-latency long-context serving prioritizes per-request latency over aggregate throughput, deployments operate at modest batch sizes of 1 to 32~\cite{zhong2024distserve, sarathi}. A larger batch raises decode latency, since each request attends only to its own KV cache and per-request attention cannot be fused.}

% while user experience remains highly sensitive to decode latency, often measured as time per output token~\cite{distserve}. Batch size is therefore often kept modest to meet latency targets, which in turn makes both the projection and core attention kernels memory-bound. Low-latency, long-context attention serving is thus an important systems task.

% \subsection{GPU Parallelism for Attention}

\subsection{Processing In- and Near-Memory}
\begin{figure}[t]
    \centering
    \includegraphics[width=\columnwidth]{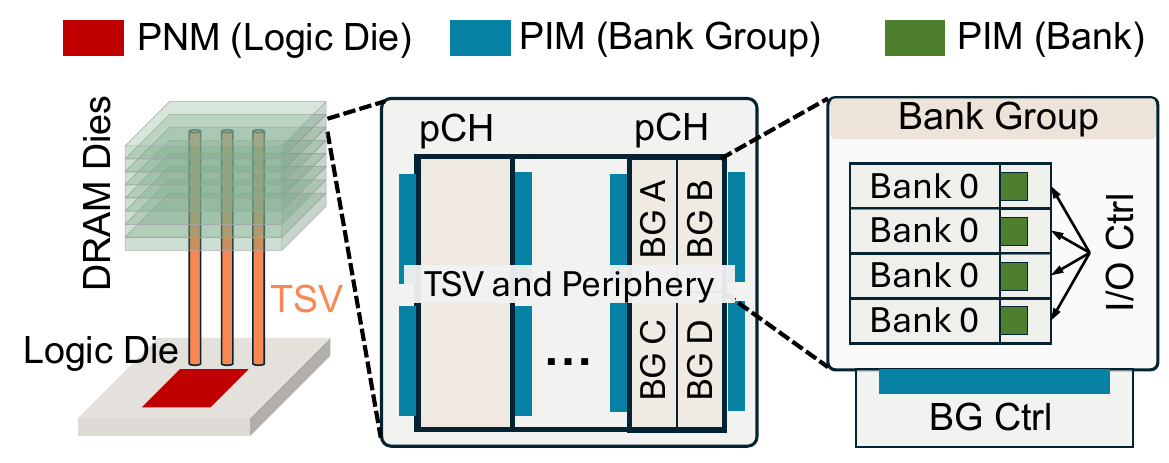}
    \caption{PIM/PNM architecture overview.}
    \label{fig:nmp}
\end{figure}

As shown in \autoref{fig:nmp}, compute units can sit at two places in an HBM stack, each with distinct trade-offs.

\textbf{Processing in memory (PIM)} places arithmetic units
directly on DRAM dies~\cite{hbm_pim, aim-gddr6, attacc, neupims}.
By co-locating compute with individual banks or bank groups,
PIM bypasses the TSV bus bottleneck and exposes full DRAM array
bandwidth, an order-of-magnitude increase over the external HBM
interface.
However, compute units consume cell area, directly reducing
storage capacity.
Moreover, DRAM dies are manufactured on mature nodes optimized
for cell density (e.g., 1$\beta$, 1$\gamma$, 1c) rather than
logic performance, and the tight area and power envelope
restricts in-DRAM compute to simple GEMV operations.
Commercial products such as Samsung
HBM2-PIM~\cite{hbm_pim} and SK Hynix AiM~\cite{aim-gddr6} adopt
this approach, but their limited compute capability and reduced
capacity make them insufficient for modern attention workloads.

\textbf{Processing near memory (PNM)} integrates compute units into the logic die~\cite{li2025h2llm, niu2022184hb_nmp}, incurring no capacity penalty. Because the logic-die process node is independent of the DRAM array, PNM permits far more complex logic than PIM, especially from HBM4 onward where logic dies move to $\le$5\,nm~\cite{hbm4_spec}, providing sufficient transistor density for sophisticated logic. 
PNM alone does not increase per-cube bandwidth, but replacing GPU compute dies with additional PNM-equipped cubes roughly doubles the aggregate package bandwidth. 
Area and power, however, remain binding constraints: the logic die already hosts the memory controller and PHY, and sits beneath DRAM arrays that dissipate substantial heat. Excessive power risks thermal violations and DRAM errors, so the added compute must be carefully budgeted, motivating the microarchitecture choices in \autoref{sec:arch_overview}.

% We are standing at a turning point where NMP is no longer a fairy tale
% but a practical architecture. With 2\,nm--5\,nm technology, it
% is feasible to integrate a complete accelerator on the HBM logic
% die. \xname{} represents a fundamental departure from prior IMC
% and NMP designs: rather than serving as an accessory to a GPU or
% NPU, it is an independent, fully programmable accelerator
% tailored for low-latency, memory-bound tasks. We hope this work
% inspires further research on memory-centric architectures, which
% show great potential for such workloads.

\section{Motivation}

\subsection{Limitation of GPUs}
% Add GPU LLC hit rate figure.

\begin{figure}[t]
    \centering
    \includegraphics[width=\columnwidth]{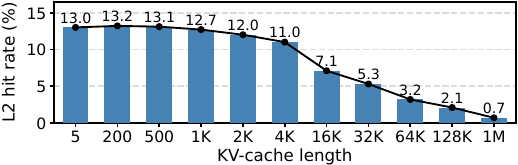}
    \caption{LLC hit rate for Qwen3-235B at batch 16.}
    \label{fig:LLC}
\end{figure}

%todo: 
%1: increase batch size to 16 , 2: increase #seq_len, 3:caption 
%double check batch 大小
%1. 完整的QWEN模型 2.Batch是否拉到了16 

\begin{figure}[t]
    \centering
    \includegraphics[width=0.49\textwidth]{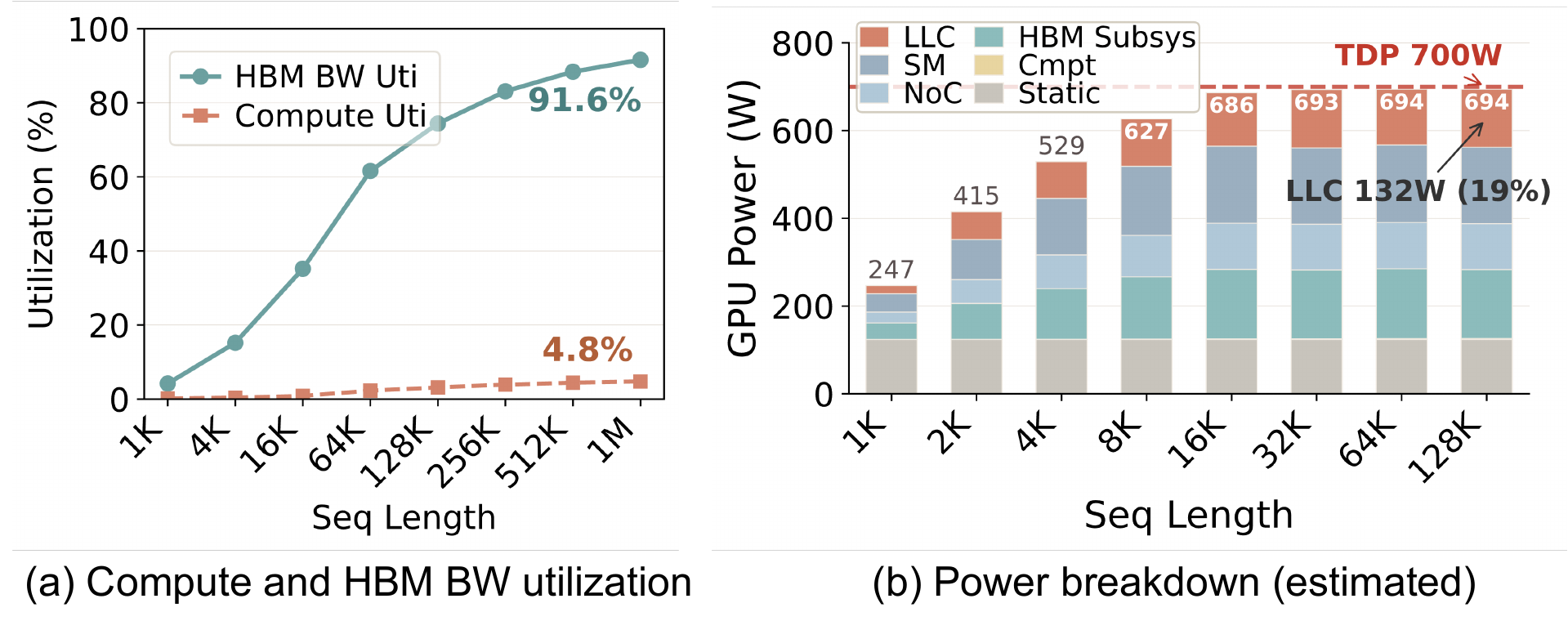}
    \caption{Profiling of H100 (a) hardware utilization and (b) power consumption for Qwen3-235B attention (batch=1).}
    \label{fig:3_profile}
\end{figure}

\color{black}
Most existing LLM serving systems rely on GPUs for decode
attention. However, GPUs are architected for compute throughput,
making them a poor fit for attention, which is memory-bound.
As shown in~\autoref{fig:3_profile}(a), HBM bandwidth is
nearly saturated at over 90\% utilization while GPU compute
units remain largely idle at under 5\%, representing significant
area and power waste.

\begin{figure}[t]
    \centering
    \includegraphics[width=0.44\textwidth]{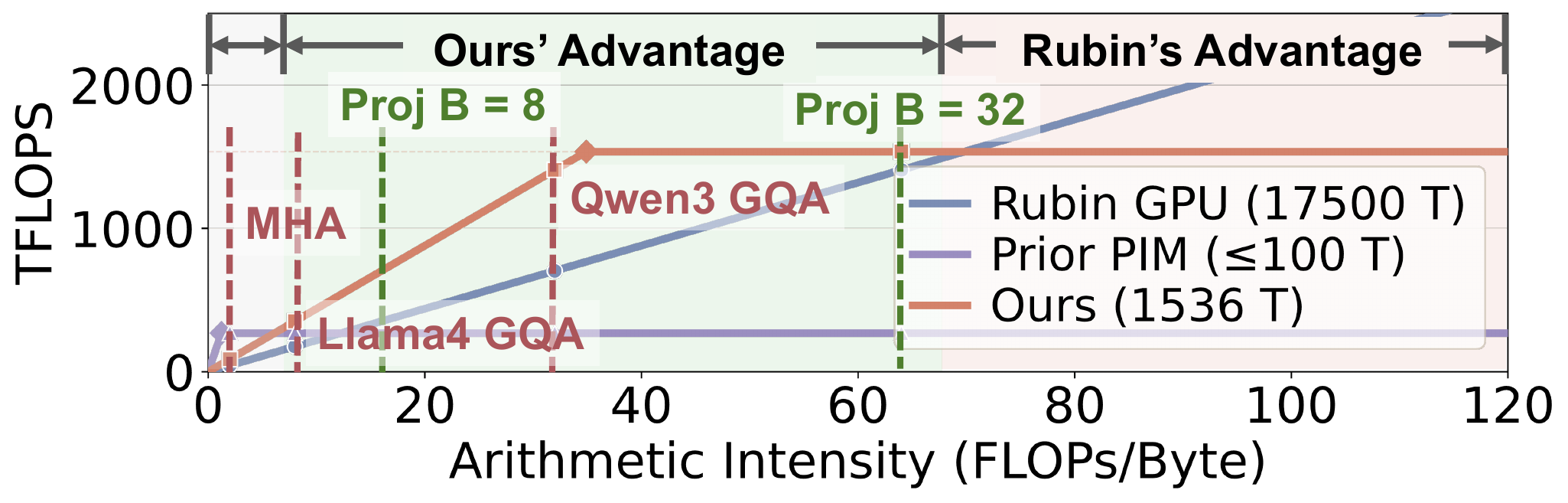}
    \caption{A roofline analysis of Rubin and \xname{}.}
    \label{fig:3_roofline}
\end{figure}

The roofline analysis in~\autoref{fig:3_roofline} confirms
this mismatch. The arithmetic intensity of the attention kernel
falls far below the GPU's compute-to-bandwidth ratio, placing it in the memory-bound regime where additional compute
yields no performance benefit.
Motivated by this observation, \xname{} replaces GPU compute dies with HBM-PNM cubes to provide substantially higher aggregated memory bandwidth, demonstrating strong potential to accelerate reasoning and agentic
workloads that require extremely long contexts.
% a memory-centric architecture that replaces GPU compute dies with
% HBM-PNM cubes to provide substantially higher memory bandwidth to match attention demands.
% Although~\xname{} has lower peak compute throughput than Rubin
% and is therefore less suited to compute-intensive kernels, it
% demonstrates strong potential to accelerate emerging reasoning and agentic
% workloads that require extremely long contexts.

\subsection{Power Breakdown of GPU}
% As AI compute demand outpaces energy infrastructure growth,
% power consumption has become a first-order data center
% constraint.
To understand how GPUs spend their power budget, we profile
an H100 running Qwen3-235B~\cite{qwen3} attention at batch
size 1, using NVML~\cite{nvidia_nvml} for measurement and
CACTI~\cite{muralimanohar2009cacti} for modeling.
\autoref{fig:3_profile}(b) reveals two counter-intuitive
observations.

\textbf{GPUs reach TDP even on memory-bound workloads.}
Long-context attention uses only 5\% of compute, yet measured power averages 693\,W, within 1\% of H100's 700\,W TDP.
With static power at 125\,W and total HBM power under 200\,W,
roughly 370\,W remains unaccounted for.
Our breakdown traces this to the NoC, LLC, and non-compute SM
logic, whose combined consumption grows with context length as
more SMs are activated for parallelism.
Their compute units idle, while the surrounding data-movement and control fabric still draw power.
Because~\xname{} operates under a far tighter power envelope
than a discrete GPU, adopting a GPU-like microarchitecture is
infeasible.

\textbf{The LLC consumes significant power yet contributes
little.}
As \autoref{fig:LLC} shows, the LLC hit rate for Qwen3-235B at batch 16 stays around 13\% and drops below 1\% for longer sequences. 
The LLC nonetheless dissipates 130\,W during long-context attention, 
spending significant power while contributing almost no hits.
Register files and shared memory are likewise provisioned for data reuse that attention does not exhibit, an area and power liability.
These findings confirm that GPU microarchitectural conventions
are ill-suited to~\xname{}.
% Non-compute overheads push the chip to its thermal limit
% regardless of arithmetic utilization, while large SRAM
% structures waste power without aiding performance.
A dedicated logic-die architecture that channels its area and power budget toward fully exploiting HBM bandwidth is
therefore essential.

\color{black}

\subsection{Limitations of Prior PIM/PNM Proposals}

Prior PIM/PNM works share two fundamental limitations,
% (\autoref{tab:nmp_comparison})
making them unsuitable for modern GQA/MLA attention.

\begin{figure}[t]
    \centering
    \includegraphics[width=0.45\textwidth]{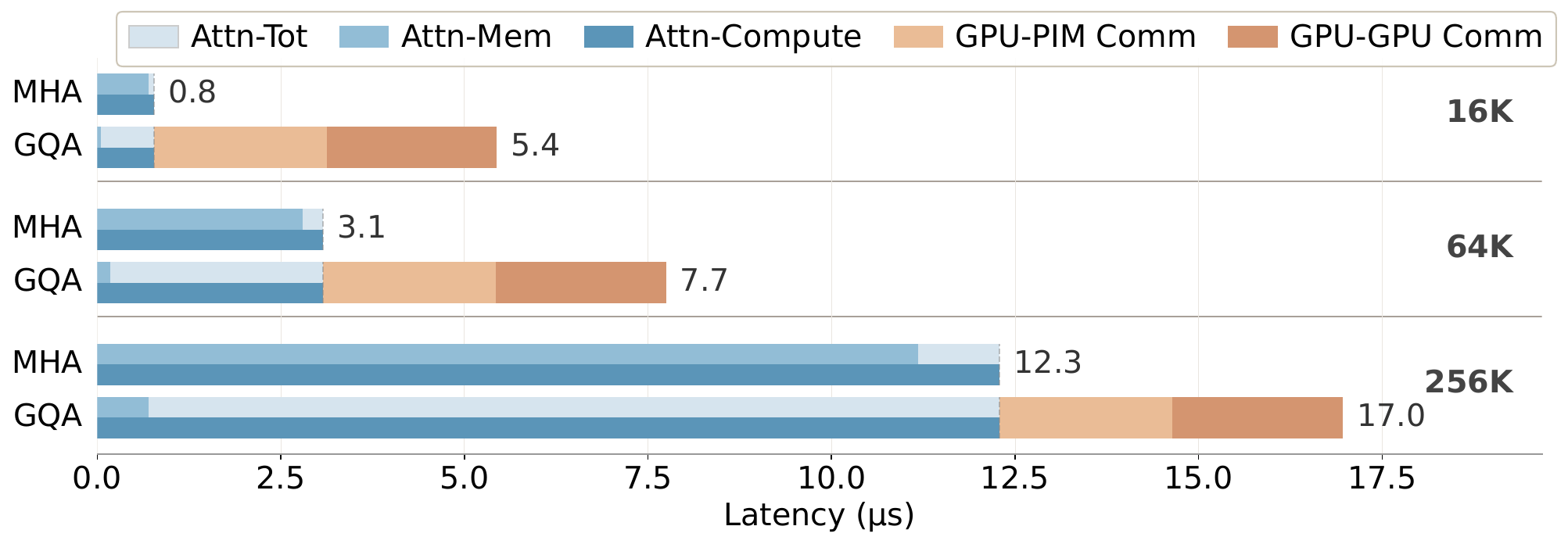}
    \caption{Qwen3-like GQA (4 KV heads) and comparable MHA (64 KV heads), both with 64 Q heads, on prior PIM. Prior PIMs rely on GPUs for collectives and have too little compute to exploit GQA's reduced traffic.}
    \label{fig:3_pim_gqa_mha}
\end{figure}

\textbf{GPU-centric paradigm introduces communication overhead.}
Existing works such as AttAcc~\cite{attacc} treat the GPU as the central device for inter-cube communication. This works well for MHA, where sufficient KV heads allow each PIM device to independently handle one head via TP. 
However, GQA and MLA compress KV heads by 16--128$\times$, so TP alone can no longer partition work across PIM devices (e.g., 64 devices vs.\ 4 KV heads), forcing context parallelism (CP) along the sequence dimension. CP introduces AllReduce, which is disastrous for GPU-centric PIM.
As shown in \autoref{fig:3_pim_gqa_mha}, AttAcc connects 8 GPUs and 64 PIM devices together via NVLink.
When handling AllReduce, each GPU must collect data from 8 PIM devices first and then perform a cross-GPU reduction. 
This communication overhead alone is 1.5$\times$ longer than the attention computation at 64K, expanding to 24$\times$ once sufficient compute is added to relieve the compute bottleneck discussed below. 
These results further assume an ideal NVLink latency of 900 ns with no kernel launch overhead, whereas our real-GPU profiling shows that a single 8 B transfer takes over 12{,}000 ns end-to-end.

\xname{} eliminates this bottleneck by departing from the GPU-centric paradigm. Each HBM cube carries its own logic die and acts as a fully autonomous accelerator. Cubes interconnect through high-speed D2D links to form a standalone multi-chiplet attention processor with no host GPU.

\textbf{Insufficient compute power bottlenecks GQA.}
Most prior PIM designs place GEMV units in the DRAM die, which suffice for MHA but not for GQA. As shown in \autoref{fig:3_roofline}, GQA increases arithmetic intensity by 16--32$\times$ through KV head sharing, rendering prior architectures compute-bound. \autoref{fig:3_pim_gqa_mha} confirms that although GQA reduces memory traffic by 16$\times$ over MHA, total attention latency shows no improvement due to this bottleneck. This shift necessitates GEMM-capable units far too large for DRAM dies, driving a migration to logic-die PNM. Duplex~\cite{yun2024duplex} has explored this direction, yet its $\sim$28\,nm logic-die process severely constrains the design space. We propose \xname{} at this turning point, where advanced logic-die processes make sophisticated microarchitecture feasible and unlock new potential for memory-centric acceleration.

% \newcommand{\cmark}{\textcolor{green!60!black}{\checkmark}}
% \newcommand{\xmark}{\textcolor{red!70!black}{$\times$}}
% \begin{table}[h]
% \centering
% \caption{Comparison of representative PNM works.}
% \label{tab:nmp_comparison}
% % \small
% \setlength{\tabcolsep}{4pt}
% \resizebox{0.95\columnwidth}{!}{%
% \begin{tabular}{lcccccc}
% \toprule
%  & DRAM & Cmpt & Cmpt Unit & Cmpt Unit & GPU & Multi-chiplet \\
%  & tech & unit & location & process & free & connection \\
% \midrule
% NDPBridge~\cite{tian2024ndpbridge} & DDR4     & GEMV & Near-DIMM & $\sim$28\,nm          & \xmark & \xmark \\
% AttAcc~\cite{attacc}               & HBM2/3      & GEMV & DRAM die  & 1$\alpha$ ($\sim$28--40\,nm) & \xmark & \xmark \\
% NeuPIMs~\cite{neupims}             & HBM2/3      & GEMV & DRAM die  & 1$\alpha$ ($\sim$28--40\,nm) & \xmark & \xmark \\
% ORCHES~\cite{li2025orches}         & HBM-like & GEMV & DRAM die  & 1$\alpha$ ($\sim$28--40\,nm) & \xmark & \xmark \\
% Duplex~\cite{yun2024duplex}        & HBM3      & GEMM & Logic die & $\sim$28\,nm   & \xmark & \xmark \\
% % Stratum~\cite{stratum}             & HBM-like & GEMV & In-DRAM   & 1$\alpha$ ($\sim$28--40\,nm logic-eq.) & \xmark & \xmark \\
% \midrule
% \textbf{\xname{}}  & \textbf{HBM4+} & \textbf{GEMM} & \textbf{Logic die} & $\leq$\textbf{5\,nm} & \cmark & \cmark \\
% \bottomrule
% \end{tabular}}
% \end{table}

\section{Architecture}

\subsection{Overview}
\label{sec:arch_overview}

\begin{figure*}[t]
      \centering
      \includegraphics[width=0.93\textwidth]{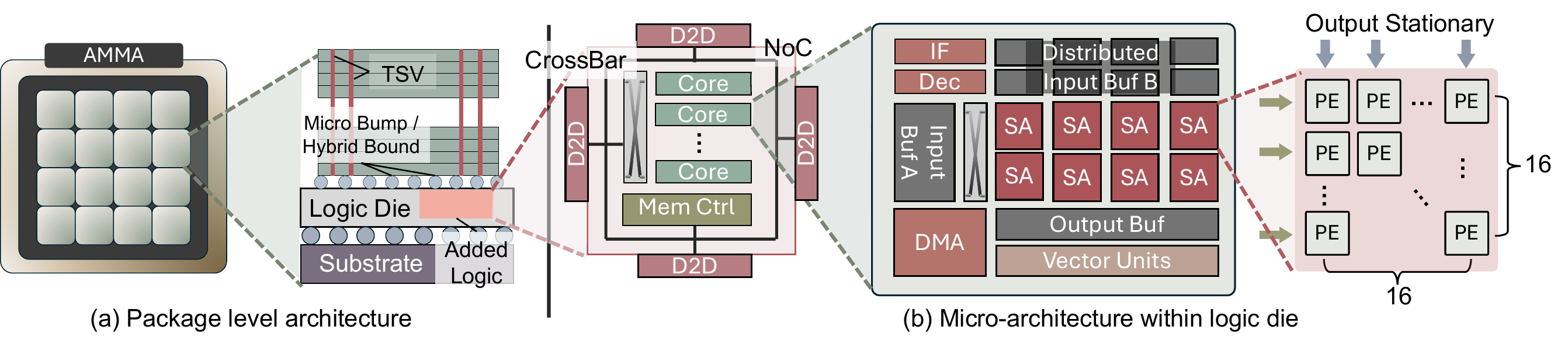}
      \caption{\xname{} architecture hierarchy.
               (a)~Package and cube level integration and
               (b)~Core and SA level microarchitecture.}
      \label{fig:arch}
\end{figure*}

The \xname{} architecture is illustrated in~\autoref{fig:arch}.
Each chip package arranges 16 HBM-NMP cubes in a $4{\times}4$
2D mesh, where each cube integrates an HBM stack with a
PNM-enabled logic die and communicates through high-speed D2D
links.
Because attention workloads differ fundamentally from
conventional GPU workloads in access patterns and resource
balance, \xname{} departs from GPU-style designs and adopts a
microarchitecture guided by three principles.

\textbf{(P1) Many small SAs over few large ones.}
Decode attention's $M$ dimension is tiny (1--32, set by batch size or
Q-heads per KV-head in GQA), so a TPU-style $128{\times}128$ array
would leave $>$87\% of PEs idle.
We instead deploy 96 $16{\times}16$ systolic arrays, sized to match
the workload's narrow $M$ and keep every PE fully occupied while
delivering the same aggregate throughput.
We choose systolic arrays over vector units because SAs reuse each operand across 16 PEs,
reducing the required SRAM read bandwidth by $16{\times}$ and thus
significantly lowering area and power overhead.
% {

% GPUs also contain many small $16{\times}16$-scale arrays inside their tensor
% cores, so the distinction is not array size but how operands are supplied to
% the arrays. A GPU feeds these arrays through a deep register-file and LLC
% hierarchy built to manufacture reuse, which decode attention does not have.
% \xname{} instead streams operands directly from HBM into the arrays and relies
% on the systolic reuse above to lower SRAM read bandwidth, reaching
% tensor-core-class throughput without that cache hierarchy and its area and
% power cost.}

\textbf{(P2) LLC-free architecture.}
An LLC is unnecessary from both the architecture and workload
perspectives.
From the architecture side, a GPU LLC amplifies effective
bandwidth to bridge the gap between compute and memory bandwidth. \xname{}, however, sizes compute to match raw HBM bandwidth, naturally eliminating this gap.
From the workload side, an LLC harnesses data reuse to reduce
redundant DRAM accesses, yet in low-latency decode attention
with small batch sizes ($B{=}1$--$32$), the KV cache and input
requests stream through only once, leaving little reuse to
capture.
Removing the LLC reclaims 20\% of the power budget and
substantial die area, which we redirect to useful compute.

\textbf{(P3) Two-level crossbar for data sharing.}
Removing the LLC also removes its data-exchange role
among cores. 
We replace it with a two-level crossbar that broadcasts shared inputs and collects partial outputs across the 96 SAs.
On the input side, the same query vector is needed by all SAs
working on different tiles, and the crossbar delivers it
without redundant DRAM reads.
On the output side, partial sums from multiple SAs are collected through the crossbar, avoiding costly DRAM traffic.
We organize the crossbar hierarchically, with a local crossbar
within each 8-SA core and a global crossbar across 12 cores,
reducing area from $O(N^2)$ to $O(N\!\sqrt{N})$
compared with a flat design over all SAs.

\subsection{Microarchitecture Inside Cubes and Cores}
\label{sec:arch_detail}

The 16 cubes form a 2D mesh, with each cube connected to up to
four neighbors via die-to-die (D2D) links.
{
We adopt a mesh because each D2D link connects only adjacent dies, which minimizes wire length, making it more practical for manufacturing and eliminating the need for error correction to achieve shorter latency.}
Within each cube, as shown in \autoref{fig:arch}(a), DRAM dies
sit atop the logic die and are connected through micro bumps or hybrid-bonding.
The logic die hosts the HBM PHY and memory controller in its
standard region, while our compute logic is added to enable PNM.

As shown in \autoref{fig:arch}(b), each cube contains 12
compute cores, a memory controller, and four D2D ports, all
connected through NoC.
The 12 cores are further interconnected via the two-level
crossbar described in P3, enabling query broadcast across cores
computing different $N$-tiles and cross-core reduction of
partial sums.
% MA engines stream data directly from HBM
% into per-core buffers, and the NoC is sized to sustain
% concurrent computation and inter-cube communication without
% contention.

Each core has an instruction front-end, a DMA engine that streams directly from HBM into per-core buffers, two input-buffer banks, a cluster of 8 SAs, an output buffer, and a vector unit.
\emph{Input Buf~A} feeds each $A$-matrix row slice to its SA, or broadcasts shared rows to all 8 SAs via the local crossbar when reuse arises.
\emph{Distributed Input Buf~B} is partitioned across the SA
cluster so that each SA receives its own $B$-matrix slice with
minimal wire length.
Both input buffers are double-buffered to hide the 200\,ns HBM latency, prefetching the next tile while the SAs compute the current one.
Each core totals 32\,KB of SRAM ($2{\times}6.4$\,KB for
Input Buf~A, $2{\times}6.4$\,KB for Distributed Input Buf~B,
and 6.4\,KB for the Output Buf), yielding 3\,MB per cube and
48\,MB across all 16 cubes.
This is over $2{\times}$ smaller than H100 and over $6{\times}$ smaller than Rubin, with proportional power and area savings.
The vector unit handles post-SA element-wise operations such as softmax and linear norm, reading the output buffer in a pipeline that overlaps the next tile's SA computation.

\begin{figure}[t]
      \centering
      \includegraphics[width=0.46\textwidth]{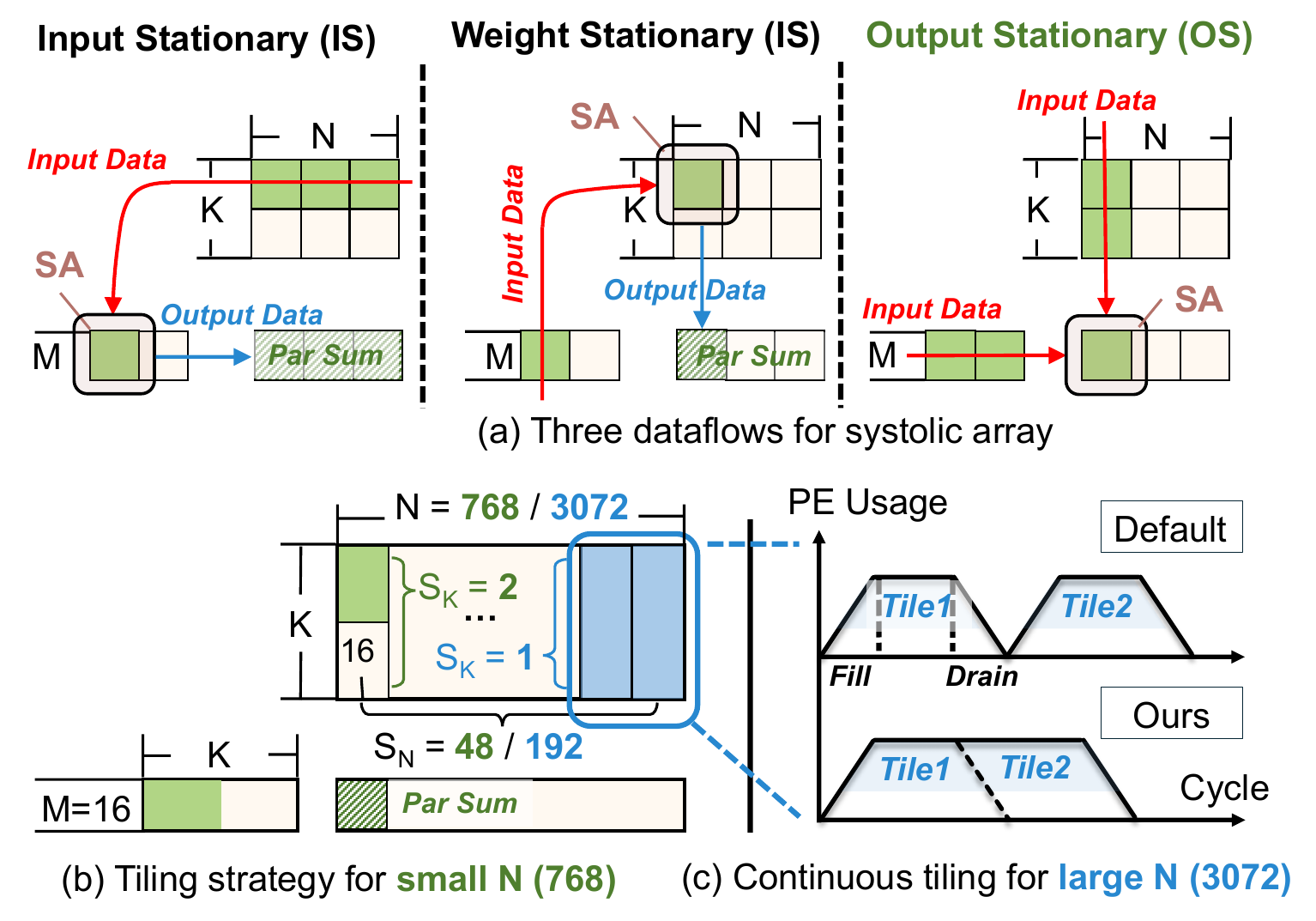}
      \caption{(a)~Three conventional SA dataflows: WS, OS, IS.
               (b)~Tiling strategy for small $N$ and large $N$.
               (c)~Continuous tiling for large $N$: fill of tile $i{+}1$ overlaps drain of
               tile $i$.}
      \label{fig:arch_tiling}
\end{figure}

\subsection{Systolic Array Dataflow}
\label{sec:arch_dataflow}

\xname{} deploys 96 small $16{\times}16$ SAs operating in parallel
with a small $M$ dimension ($M{=}1$--16, set by batch size or the
number of Q-heads per KV-head in GQA).
TPU-style architectures pair a few large $128{\times}128$ SAs with
weight-stationary dataflow, relying on large-batch workloads to
supply a large $M$ that keeps every PE busy.
Many small arrays with a small $M$ demand a different dataflow
and tiling strategy.

As illustrated in \autoref{fig:arch_tiling}(a), each canonical SA dataflow keeps a different operand stationary and
streams one GEMM dimension through the array, which must be
large for high utilization: \textbf{WS}
streams $M$, \textbf{IS} streams $N$, and \textbf{OS} streams $K$.

We choose OS for \xname{} because it is the only dataflow that
suits our long-context, small-$M$ regime.
WS is ruled out immediately since streaming only $M{=}1$--$16$
rows through a $16{\times}16$ weight tile wastes the majority of
PE cycles.
IS appears attractive because $N$, the sequence length, is
inherently large, but under IS each SA accumulates a partial output
that grows with $N$, and collecting these partials across 96
parallel SAs incurs $O(N)$ communication that worsens linearly as
sequence length increases.
OS avoids both problems, as $K$ is large enough for streaming and every completed output tile is a fixed $16{\times}16$ block,
making the cross-SA collection cost independent of sequence length.

\begin{figure*}[t]
      \centering
      \includegraphics[width=0.8\textwidth]{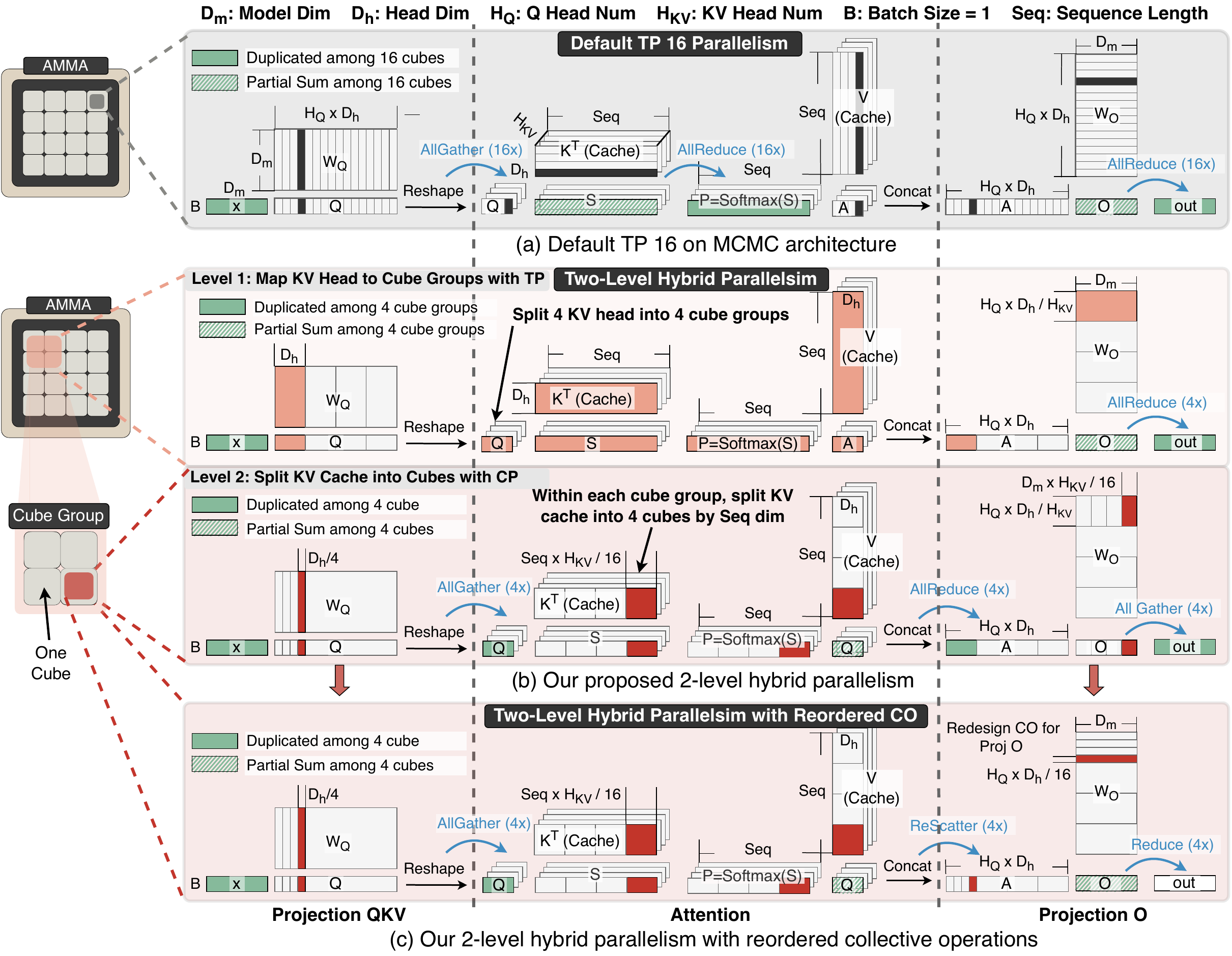}
      \caption{Parallelism design for \xname{}.
               (a)~Na\"ive TP16 distributes every attention stage across all 16 cubes.
               (b)~Our two-level hybrid parallelism maps KV heads to four cube groups
               via TP (Level~1), then partitions the KV cache by sequence within each
               group via CP (Level~2).
               (c)~The same hierarchy combined with the reordered collective flow
               described in \autoref{sec:collective}.}
      \label{fig:parallelism}
\end{figure*}

\subsection{Tiling Strategy for High Utilization}
\label{sec:arch_tiling}

{
With OS established, the remaining question is how to tile the
GEMM dimensions across the 96 SAs to maximize utilization.
We use two techniques.
A \emph{tiling methodology} partitions $K$ and $N$ to avoid idle SAs, and \emph{continuous tiling} pipelines the consecutive
tiles assigned to each array to remove inter-tile idle cycles.

Consider a GEMM $\mathbf{C} = \mathbf{A}\mathbf{B}$ with shapes $M{\times}K$
and $K{\times}N$.
We tile $N$ into $N/16$ column tiles and optionally tile $K$ into $S_K$
segments of depth $k {=} K/S_K$, yielding $T {=} S_K {\cdot} N/16$ output
tiles of size $16{\times}16$.
Let $P{=}96$ denote the number of available SAs, and let
$n {=} \lceil T/P \rceil$ denote the number of consecutive tiles that each
busy array processes.
Overall utilization decomposes into two factors:
\begin{equation}
  U_\text{total}
  = \underbrace{\frac{\min(T,\,P)}{P}}_{\text{fraction of SAs busy}}
  \;\times\;
  \underbrace{\frac{n\,k}{n\,k + 2(M_\text{SA} {-} 1)}}_{\text{per-SA efficiency}},
  \qquad n = \left\lceil T/P \right\rceil .
  \label{eq:util}
\end{equation}
The first factor is the fraction of the 96 arrays that are busy.
The second is the per-array efficiency, where an array spends $nk$ cycles on
useful work and pays the fill and drain overhead of $2(M_\text{SA}{-}1){=}30$
cycles, fifteen at each end, only once across its $n$ tiles.
When an array holds a single tile, $n{=}1$ and the second factor reduces to
the per-tile form $k/(k{+}30)$.
}

\textbf{Tiling methodology: }
{
Our strategy follows one principle.
We split $N$ first and split $K$ only for short sequences, just enough to give every array a tile.
Below that point, splitting activates idle arrays and raises utilization.
Beyond it, splitting brings no further gain and only adds reduction cost.
We now derive both halves of this principle from \autoref{eq:util}.

When the sequence is short, $T {\leq} P$ and $n{=}1$, and SAs stay idle if we split only $N$. We then simplify \autoref{eq:util} by substituting $T {=} (K/k)(N/16)$:
\begin{equation}
  U_\text{total}\big|_{T \leq P}
  = \frac{(N/16) \cdot K}{P\,\bigl(k + 2(M_\text{SA}{-}1)\bigr)}
  = \frac{\text{const}}{k + 30}.
  \label{eq:util_expand}
\end{equation}
The utilization increases as we have smaller tile size $k$.
This is because splitting $K$ creates more tiles and activates idle arrays, and this parallelism gain outweighs the per-array efficiency loss.
Splitting $K$ is thus the right lever whenever arrays are underutilized.
Once every array is busy, \autoref{eq:util} falls into the $T {\geq} P$ regime. The first factor locks
to 1, and continuous tiling sets the per-array factor to $nk/(nk{+}30)$.
% Here $nk {=} NK/(16P)$ to first order, since $n {\approx} T/P {=} S_K(N/16)/P$
% and $k {=} K/S_K$ cancel the split factor $S_K$.
% The useful work per array is therefore fixed by the problem size and does not
% depend on how finely $K$ is split.
Further splitting yields no utilization gain and adds cross-array
partial-sum traffic because each output tile must then sum $S_K$
partials.
}

We illustrate with two cases ($P{=}96$) in \autoref{fig:arch_tiling}(b).
When $N{=}768$, there are only $768/16{=}48$ column tiles,
fewer than the 96 SAs. Setting $S_K{=}2$ doubles the tile count
to $T{=}96$, giving every SA exactly one tile while keeping the
per-tile depth $k{=}K/2$ well above $M_\text{SA}$.
When $N{=}3072$, there are $3072/16{=}192$ column tiles,
already exceeding 96. No $K$-splitting is needed ($S_K{=}1$),
and each SA processes two consecutive tiles at full depth
$k{=}K$.

\textbf{Continuous tiling: }
{
The amortization in \autoref{eq:util} relies on a systolic-array
pipelining technique that we adopt here.
When an array handles multiple consecutive tiles, the default schedule
serializes the drain of one tile and the fill of the next, leaving
progressively freed PEs idle during the 15-cycle drain phase, shown as
Default in \autoref{fig:arch_tiling}(c).
Continuous tiling removes this waste by feeding the next tile into the PEs
that the current tile has just released.
As tile $i$ drains from one end of the array, tile $i{+}1$ fills from the
other, so the two phases fully overlap.
Double-buffered DMA prefetch keeps operands ready before each drain.

With this overlap, an array that processes $n$ consecutive tiles pays the
$2(M_\text{SA}{-}1)$ fill and drain cycles once for the whole run rather than
once per tile, which is the per-array factor of \autoref{eq:util}:
\begin{equation}
  U_\text{SA}^\text{cont}(k, n)
  = \frac{nk}{nk + 2(M_\text{SA} {-} 1)}
  \;\xrightarrow{n \to \infty}\; 1.
  \label{eq:cont_util}
\end{equation}
For $N{=}3072$ with $n{=}2$ tiles per array, continuous tiling raises
efficiency from 52\% to 67\% at $k{=}32$, and to 81\% at $n{=}4$.
% In longer contexts where each array handles tens of tiles, every array
% approaches full utilization.
}

\section{Intra-package Parallelism}\label{sec:parallelism}

The previous section described the compute workflow within each cube.
We now distribute the full attention pipeline across all 16 HBM-NMP cubes within a single chip.
% This requires reconciling two opposing pressures: all 16 cubes provide
% massive aggregate memory bandwidth that encourages wide distribution of
% weights and the KV cache, but the 2D-mesh interconnect introduces both
% latency and bandwidth overhead for every inter-cube transfer.

\subsection{Default GPU-like Parallelism}
Existing multi-GPU systems face a similar design question, but
their different constraints and goals make their solutions
inapplicable to \xname{}.
GPU clusters keep the TP degree below the KV-head count and use data parallelism (DP) to scale throughput across more than $10{,}000$ requests, so each TP group covers a full KV head. Splitting a head across TP devices adds an AllReduce after every attention stage and dominates latency at large TP degrees.

Neither DP nor TP transfers to \xname{}.
The design goal of \xname{} is low latency for a small number of requests (1--32), not
aggregate throughput across many concurrent requests, so DP provides no benefit.
All 16 cubes must therefore cooperate on the same attention request
simultaneously to meet the low-latency goal, making it essential to distribute both the weights and the KV cache across all cubes efficiently.
Na\"ively applying TP across all 16 cubes
(\autoref{fig:parallelism}(a)) triggers 16-way AllGather and AllReduce
operations at every attention stage, stretching communication across the
full mesh diameter.
Worse, TP16 makes the transmitted data volume proportional to the \emph{sequence} dimension, drastically increasing latency at 1M context length.

% The key observation is that attention exposes two forms of parallelism with
% very different communication characteristics.
% The first is \emph{head-level independence}: different KV heads are
% mathematically separate until the final output projection, so partitioning
% along heads introduces minimal coupling.
% The second is \emph{sequence-level partitioning}: the KV cache can be split
% along the sequence dimension, with each partition contributing a partial
% attention result that is later combined.
% Rather than treating these two dimensions uniformly under a single TP degree,
% we map them hierarchically to the physical topology of the package.

\subsection{Two-Level Hybrid Parallelism}

To address this, we propose a two-level hybrid parallelism scheme illustrated in \autoref{fig:parallelism}(b).
We demonstrate with an MHA configuration of 4 KV heads and 4 Q heads for clarity,
but the design generalizes to arbitrary head counts and attention mechanisms such as GQA.
We first divide the 16 cubes into cube groups based on the number of KV heads.
With 4 KV heads this yields 4 \emph{cube groups} of 4 cubes each, where each group is shaped as a $2{\times}2$ sub-mesh rather than a $1{\times}4$ strip to minimize the maximum hop count within a group and better support collective operations.

\textbf{Level~1: map KV heads to cube groups with TP.}
Each cube group is assigned one KV head and stores the corresponding
projection weights and KV-cache entries.
Queries are partitioned accordingly, with each group handling
all Q heads associated with its assigned KV head.
The four groups then run in parallel to cover all heads.
Because KV heads are independent during the QKV projection and
attention stages, this assignment avoids cross-group
communication during the dominant KV-cache reads, leaving a
single AllReduce across cube groups at the end of the output
projection.

\textbf{Level~2: split the KV cache within each group with CP.}
Inside each four-cube group, the cubes cooperate to compute
attention for one KV head.
To fully exploit the aggregated HBM bandwidth, we split the
KV cache along the sequence dimension across the four cubes.
During decode, the query is broadcast within the group, each
cube computes attention over its local sequence shard, and the
group combines partial results to form the final per-head
output.
This introduces one AllGather and one AllReduce, but unlike
the 16-way collectives in na\"ive TP16, both are confined to
the four-cube sub-mesh with communication volume independent
of sequence length, greatly reducing overhead.

Together, the two levels form a TP+CP hybrid that confines the
bulk of data movement to local neighborhoods and ensures no
cross-cube communication scales with the sequence length, while
still engaging all 16 cubes for aggregate bandwidth.
The TP+CP hybrid is common in training but rare at inference, where CP spans NVLink-connected chips whose latency is impractical for single-request serving. On \xname{}, cubes in a group share a package and communicate over low-latency, high-bandwidth D2D links, making CP practical at inference.

% Once this hierarchical mapping is in place, the remaining performance
% bottleneck is not \emph{whether} collectives are local, but \emph{how many}
% collectives remain on the critical path.
% Section~\ref{sec:collective} addresses this with a reordered collective flow.

\begin{figure*}[t]
  \centering
  \includegraphics[width=0.8\textwidth]{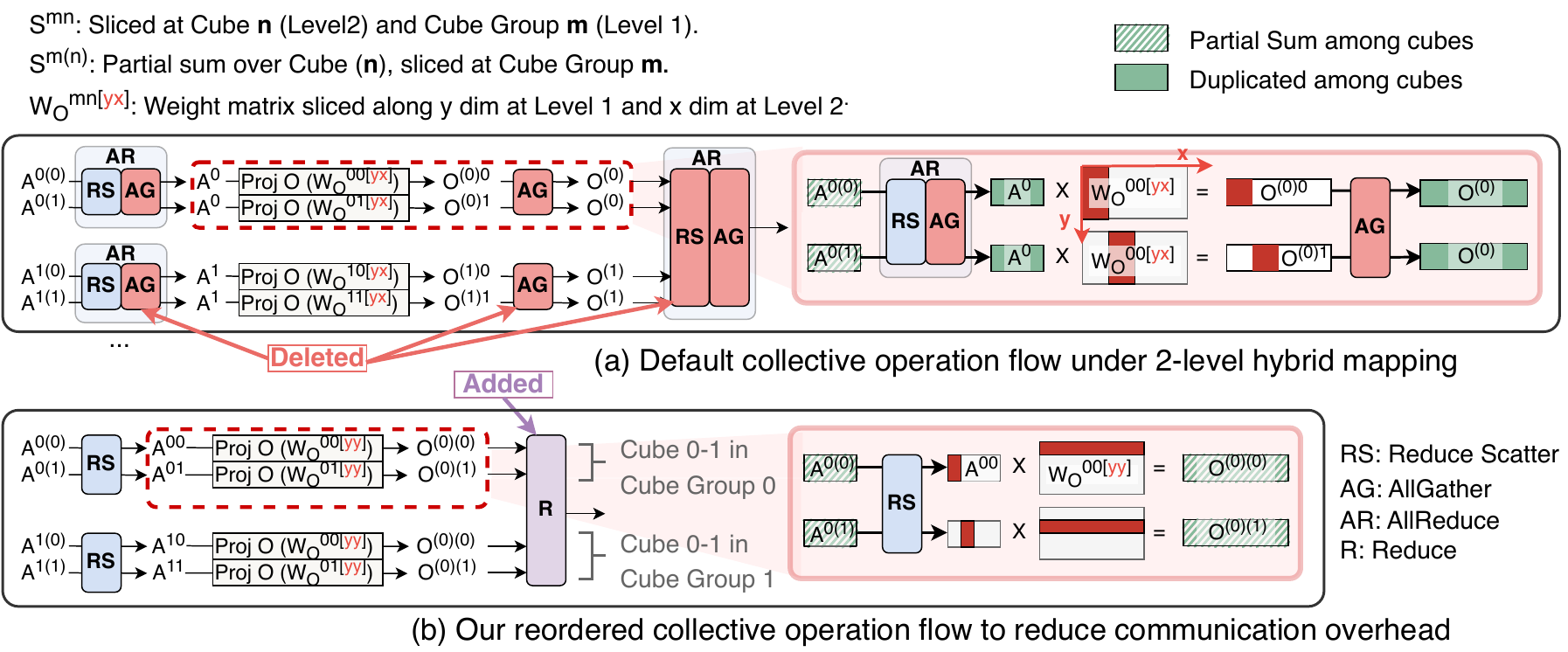}
  \caption{Reordered collective operation flow for proj O.
  (a)~Default two-level hybrid flow.
  (b)~Our reordered flow.}
  \label{fig:collective}
\end{figure*}

\section{Collective Communication Optimization}\label{sec:collective}

The two-level hybrid parallelism of \autoref{sec:parallelism} confines every collective to local neighbors and removes sequence-length-dependent traffic. We further cut the collectives on the critical path by redesigning the flow around output projection, previewed on the same two-level hierarchy in \autoref{fig:parallelism}(c).

\subsection{Reordered Collective Flow}
The default and our reordered collective flows for the two-level hybrid parallelism are illustrated in \autoref{fig:collective}.
We use the following notation throughout this section.
A superscript~$m$ denotes the cube group (Level-1 index) and
$n$ the cube within that group (Level-2 index).
$S^{mn}$ is a tensor sliced at both levels, while $S^{m(n)}$
indicates a partial sum over cube~$n$ that is sliced only at
the group level.
For the weight matrix, $W_O^{mn[yx]}$ means the Level-1
partition is along the $y$ (input) dimension and the Level-2
partition is along the $x$ (output) dimension.
In the figure, hatched blocks represent partial sums that
require reduction across cubes, and solid blocks represent
data duplicated across cubes.

The default collective flow is demonstrated in \autoref{fig:collective}(a).
After the core attention computation, each cube~$n$ in group~$m$ holds a
partial-sum activation $A^{m(n)}$ that must be reduced across its group.
The default approach applies a full intra-group AllReduce (ReduceScatter
followed by AllGather) to reconstruct the complete attention output~$A^m$ on
every cube.
Each cube then multiplies~$A^m$ by its local weight shard
$W_O^{mn[yx]}$.
Finally, a second AllReduce across groups accumulates the per-head results into the final output.

This default flow contains two sources of redundancy.
First, the intra-group AllGather replicates the full attention
output onto every cube, yet the subsequent cross-group
collective immediately scatters and re-reduces these outputs.
Second, the cross-group AllReduce broadcasts the final result
to every cube in the package, even though \xname{} operates
under a disaggregated serving model (\autoref{fig:1_intro_disaggregation})
where only the destination cube forwards the result to the remote FFN accelerator.

We address both inefficiencies with two coordinated changes, illustrated in
\autoref{fig:collective}(b).
For the first, we replace the intra-group AllReduce with a
ReduceScatter alone, so each cube~$n$ retains only a distinct
scatter slice~$A^{mn}$ rather than the full output.
To match this narrower input, we reslice the projection weight
from $W_O^{mn[yx]}$ to $W_O^{mn[yy]}$, switching the Level-2
partition axis from the output dimension to the input dimension.
Each cube then computes $A^{mn} W_O^{mn[yy]}$ independently,
producing a partial-sum output slice~$O^{(m)(n)}$ with no
weight duplication, which also eliminates the post-projection AllGather required in the default flow.
For the second, we replace the cross-group AllReduce with a point-to-point
Reduce to the single destination cube, saving roughly half the traffic.
Together, these changes remove two AllGather operations and downgrade one
AllReduce to a Reduce, while leaving per-cube compute and memory footprints unchanged.

\subsection{Correctness of the Reordered Flow}
The reordered flow is elegant, but one piece of the puzzle
remains: the softmax.
The attention weight for key position~$j$ is
\begin{equation}
  \mathrm{softmax}(s_{j})
  = \frac{e^{s_{j} - m}}{\ell},
  \qquad
  m = \max_{k}\, s_{k},
  \quad
  \ell = \sum_{k} e^{s_{k} - m},
  \label{eq:softmax}
\end{equation}
where $s_{j} = \mathbf{q}^\top \mathbf{k}_j / \sqrt{d}$.
Both $m$ and $\ell$ range over all key positions, yet context
parallelism slices the KV cache so that each cube sees only a
subset.
Two questions arise:
(1)~can partial softmax results from individual cubes be
combined to recover the exact global output, and
(2)~does inserting~$\mathbf{W}_O$ between the per-cube attention and the cross-cube reduction preserve correctness?

The first question has been answered by prior works.
FlashAttention~\cite{flashattention} shows from a
\emph{temporal} perspective that attention can be computed
incrementally over sequential tiles within a single device:
each tile~$n$ produces a local output~$\mathbf{a}_n$ together
with local softmax statistics ($m_n$ and $\ell_n$), and the correct global output is recovered as
\begin{equation}
  \mathbf{a} = \sum_{n} \alpha_n\,\mathbf{a}_n, \qquad
  \alpha_n = \frac{e^{m_n - m}\,\ell_n}{\ell},
  \qquad
  \ell = \sum_{n} e^{m_n - m}\,\ell_n,
  \label{eq:online_softmax}
\end{equation}
where $m = \max_n m_n$.
Ring Attention~\cite{ring_attention} extends this to the \emph{spatial} case: parallel devices each compute their tile independently and combine with the same correction.

The second question is specific to our reordered flow but
equally straightforward to resolve.
The correction factor~$\alpha_n$ in
\autoref{eq:online_softmax} is a per-query-position scalar while the output projection~$W_O$ is a linear map along feature dimension. The two commute:
\begin{equation}
  \Bigl(\sum_n \alpha_n\,\mathbf{a}_n\Bigr)\mathbf{W}_O
  \;=\;
  \sum_n \alpha_n\,\bigl(\mathbf{a}_n\,\mathbf{W}_O\bigr).
  \label{eq:defer}
\end{equation}
This means each cube can project first and reduce after, as long as the ReduceScatter performs the weighted sum in \autoref{eq:defer}
rather than a plain sum. In practice, each cube simply
piggybacks~($m_n$, $\ell_n$) alongside the projected output,
requiring no extra communication.

% In summary, our reordered flow extends the deferred-correction insight of
% FlashAttention and Ring Attention one step further: rather than correcting
% immediately after the attention stage, we defer the correction through the
% linear output projection and fuse it into the collective reductions that the
% projection stage already demands.
% This fusion is precisely what enables the removal of the intra-group AllGather
% described above, because without deferred correction the ReduceScatter-only
% path would feed incorrectly normalized inputs into the projection.

\begin{figure*}[t]
      \centering
      \includegraphics[width=0.95\textwidth]{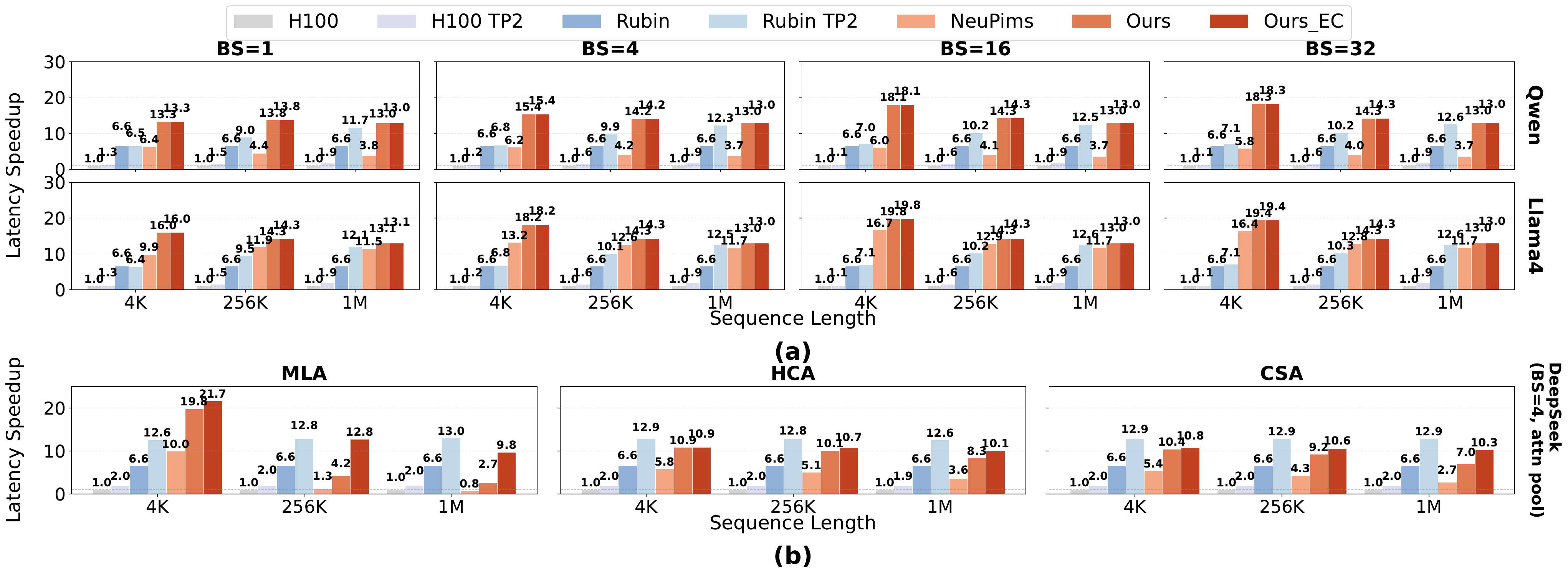}
      \caption{Decode latency speedup, normalized to H100.}
      \label{fig:fig1_speedup}
\end{figure*}
% 只保留3个seq_len, 增加Ours_EC
%double check HCA & CSA的结果(Rubin2)

%第三行增加Ds的结果
\begin{figure*}[t]
      \centering
      \includegraphics[width=0.8\textwidth]{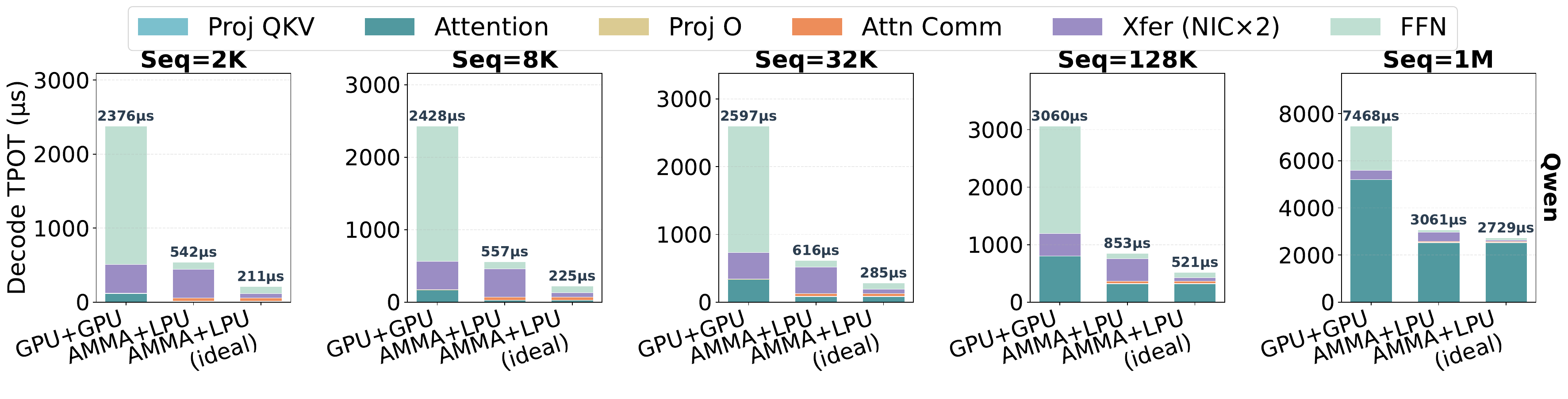}
      \caption{End-to-end decode latency breakdown.}
      \label{fig:fig1_e2e_speedup}
\end{figure*}
%figure12: 1, replace CL 2, 增加MLA
%纵向变窄拉长横向

\section{Experiments}
\textbf{Methodology.}
We combine simulation with real-GPU profiling to evaluate the performance of \xname{}.
For single-cube performance, we model each HBM-NMP cube using ScaleSim~\cite{scalesim}, an open-source systolic-array simulator.
For multi-cube performance, we use AstraSim~\cite{astrasim,astrasim2}, replacing its default GPU parameters with our HBM-NMP cube specifications and ingesting the per-cube results from ScaleSim.
For GPU baselines, we collect end-to-end latency on an 8×H100 server across a range of batch sizes and sequence lengths.
{The GPU baselines run FlashAttention-2 with FlashDecoding, which already applies context parallelism inside the kernel and represents an optimized GPU kernel.}
Because Rubin is not yet publicly available, we project its performance by scaling the H100 measurements with Rubin's published specifications while preserving the measured utilization ratios.

\textbf{Metrics.}
We use latency as the primary performance metric, as it directly reflects
user-facing serving quality.
We also report energy efficiency (Token/J) and power consumption (W),
which are of primary concern to data center operators.

\textbf{\xname{} configurations.}
\autoref{table:5_hardware_config} summarizes the \xname{} hardware parameters. We provide two versions: a standard version for GQA and an enhanced-compute (EC) version for more compute-intensive attention variants such as MLA, HCA, and CSA.
We model the inter-cube D2D latency as a fixed 15\,ns per hop, following the UCIe~3.0 protocol~\cite{ucie3.0}.
Standard \xname{} delivers 96~TFLOPS per cube from 96 $16{\times}16$ systolic arrays at 2\,GHz, an aggregate 1536~TFLOPS per chip.
The EC version uses larger $32{\times}32$ arrays to deliver $4\times$ the compute of the standard version.

% We also compare against variants of our mapping strategy, all implemented on the \xname{} architecture. We use \textbf{TP16} to denote default GPU-like tensor parallelism across HBM cubes, \textbf{HP} for our two-level hybrid mapping alone, and \textbf{HP\_RO} for the full design combining hybrid mapping with reordered scheduling.

% comprising approximately 4\,ns of adapter
% latency for protocol-to-link conversion at each endpoint, 10\,ns of PHY
% latency (Tx$+$Rx), and under 1\,ns of channel propagation delay due to
% the sub-2\,mm package reach.
% Section~\ref{sec:ablation} presents a sensitivity analysis on D2D
% bandwidth and compute capacity.

{
\textbf{Baseline configurations.}
For the multi-GPU baselines, TP$k$ denotes tensor parallelism across $k$ devices: for GQA, H100~TP2 and Rubin~TP2 split the KV heads across two chips, while for the single-KV-head MLA, HCA, and CSA they split the head dimension.}
% \textbf{Ours-EC} denotes the enhanced-compute variant of \xname{} in \autoref{table:5_hardware_config}, which replaces the base $16{\times}16$ systolic arrays with $32{\times}32$ arrays to raise per-cube compute from 96 to 384\,TFLOPS for the higher arithmetic intensity of MLA. Every \xname{} configuration uses the $4{\times}4$ on-package D2D mesh, while the GPU baselines use NVLink across chips.}
For NeuPIMs~\cite{neupims}, we scale the NPU to Rubin and the PIM to HBM4 for fairness. Its on-bank PIM provides $9\times$ higher bandwidth than the HBM interface but offers limited compute.

\textbf{Models and workloads.}
We evaluate diverse attention models: Qwen3-235B and Llama4-Maverick for GQA, DeepSeek-V3 for MLA, and DeepSeek-V4 for HCA and CSA.
For \xname{} we model QKV projection, core attention, and output projection.
We also model end-to-end latency by combining \xname{} with an LPU into a heterogeneous system that runs MoE on the LPU, driving the MoE part with open-source expert-selection traces~\cite{yu2026patternschaosforecastingdata}.
Finally, we evaluate continuous batching on realistic workloads, modeling request arrivals as a Poisson process and sequence lengths as Gaussian.
% We exclude MoE operations because both current systems and emerging industrial designs exhibit a clear trend toward attention--MoE disaggregation, where dedicated hardware such as LPUs or GPUs handles MoE while \xname{} serves as the attention engine within a heterogeneous system.

\begin{table}[]

\centering
\caption{Hardware configurations}
\label{table:5_hardware_config}
\resizebox{0.4\textwidth}{!}{
\begin{tabular}{@{}cccccc@{}}
\toprule
\multicolumn{1}{c|}{}                       & \multicolumn{1}{c|}{\textbf{H100}} & \multicolumn{1}{c|}{\textbf{Rubin}} & \multicolumn{1}{c|}{\textbf{\begin{tabular}[c]{@{}c@{}}NeuPIMs\\  GPU/PIM\end{tabular}}} & \multicolumn{1}{c|}{\textbf{\begin{tabular}[c]{@{}c@{}}Ours\\ (16 Cubes)\end{tabular}}} & \textbf{\begin{tabular}[c]{@{}c@{}}Ours-EC\\  (16 Cubes)\end{tabular}} \\ \midrule
\multicolumn{1}{c|}{Compute (FP8 TFLOPS)}   & \multicolumn{1}{c|}{1978}          & \multicolumn{1}{c|}{17500}          & \multicolumn{1}{c|}{17500/198}                                                          & \multicolumn{1}{c|}{1536}                                                               & 6144                                                                   \\
\multicolumn{1}{c|}{HBM Type}               & \multicolumn{1}{c|}{HBM3}          & \multicolumn{1}{c|}{HBM4}           & \multicolumn{1}{c|}{HBM4+PIM}                                                           & \multicolumn{1}{c|}{HBM4+PNM}                                                           & HBM4+PNM                                                               \\
\multicolumn{1}{c|}{Tot HBM BW (TB/s)}      & \multicolumn{1}{c|}{3.35}          & \multicolumn{1}{c|}{22}             & \multicolumn{1}{c|}{22/198}                                                             & \multicolumn{1}{c|}{44}                                                                 & 44                                                                     \\
\multicolumn{1}{c|}{HBM BW per cube (TB/s)} & \multicolumn{1}{c|}{0.67}          & \multicolumn{1}{c|}{2.75}           & \multicolumn{1}{c|}{2.75/24.75}                                                         & \multicolumn{1}{c|}{2.75}                                                               & 2.75                                                                   \\
\multicolumn{1}{c|}{TDP (W)}                & \multicolumn{1}{c|}{700}           & \multicolumn{1}{c|}{2200}           & \multicolumn{1}{c|}{1600/1046}                                                          & \multicolumn{1}{c|}{1440}                                                               & 1800                                                                   \\
\multicolumn{1}{c|}{Package Size (mm2)}     & \multicolumn{1}{c|}{$\sim$1516}    & \multicolumn{1}{c|}{$\sim$2568}     & \multicolumn{1}{c|}{$\sim$1516}                                                         & \multicolumn{1}{c|}{$\sim$1936}                                                         & $\sim$1936                                                             \\ \midrule
\multicolumn{6}{c}{\textbf{C2C Connection Spec}}                                                                                                                                                                                                                                                                                                                                    \\ \midrule
\multicolumn{6}{c}{C2C   NVLink: latency=900ns,  Dual direction   BW=3600 GB/s for Rubin and 900 GB/s for H100}                                                                                                                                                                                                                                                                     \\ \midrule
\multicolumn{6}{c}{\textbf{D2D Connection Spec}}                                                                                                                                                                                                                                                                                                                                    \\ \midrule
\multicolumn{6}{c}{\begin{tabular}[c]{@{}c@{}}D2D   UCIe3.0 for AMMA: latency=15ns, including \\adapter lat=4ns  (TX+RX), PHY lat=10ns (TX+RX), channel propagation delay=1ns.\\      BW=1500 GB/s\end{tabular}}                                                                                                                                                            \\ \midrule
\multicolumn{6}{c}{\textbf{Power Spec}}                                                                                                                                                                                                                                                                                                                                             \\ \midrule
\multicolumn{6}{c}{\begin{tabular}[c]{@{}c@{}}HBM3   Cube + PHY: 40W/cube, HBM4 Cube + PHY: 75W/cube,\\      H100 Compute Die: 500W/Die, Rubin Compute Die: 800W/Die,\\      AMMA Logic Die: 15W/cube and 15Wx16=240W/Chip\\      AMMA-EC Logic Die: 30W/cube and 30Wx16=480W/Chip\end{tabular}}    
                                                                                      \\ \midrule
 
\multicolumn{6}{c}{\textbf{Partial Sum Spec}}                                                                                                                                                                                                                                                                                                                                          \\ \midrule

\multicolumn{6}{c}{\begin{tabular}[c]{@{}c@{}} SA to local Vector Unit latency: 16 cycles, \\ SA to cross group Vector Unit latency: 32 cycles, \\     local Vector Unit throughput = 32 elements/cycle per SA, \\ core-level Vector Unit throughput = 256 elements/cycle\end{tabular}} 

\\ \bottomrule
\end{tabular}
}
\end{table}

\begin{figure}[t]
    \centering
    \includegraphics[width=\columnwidth]{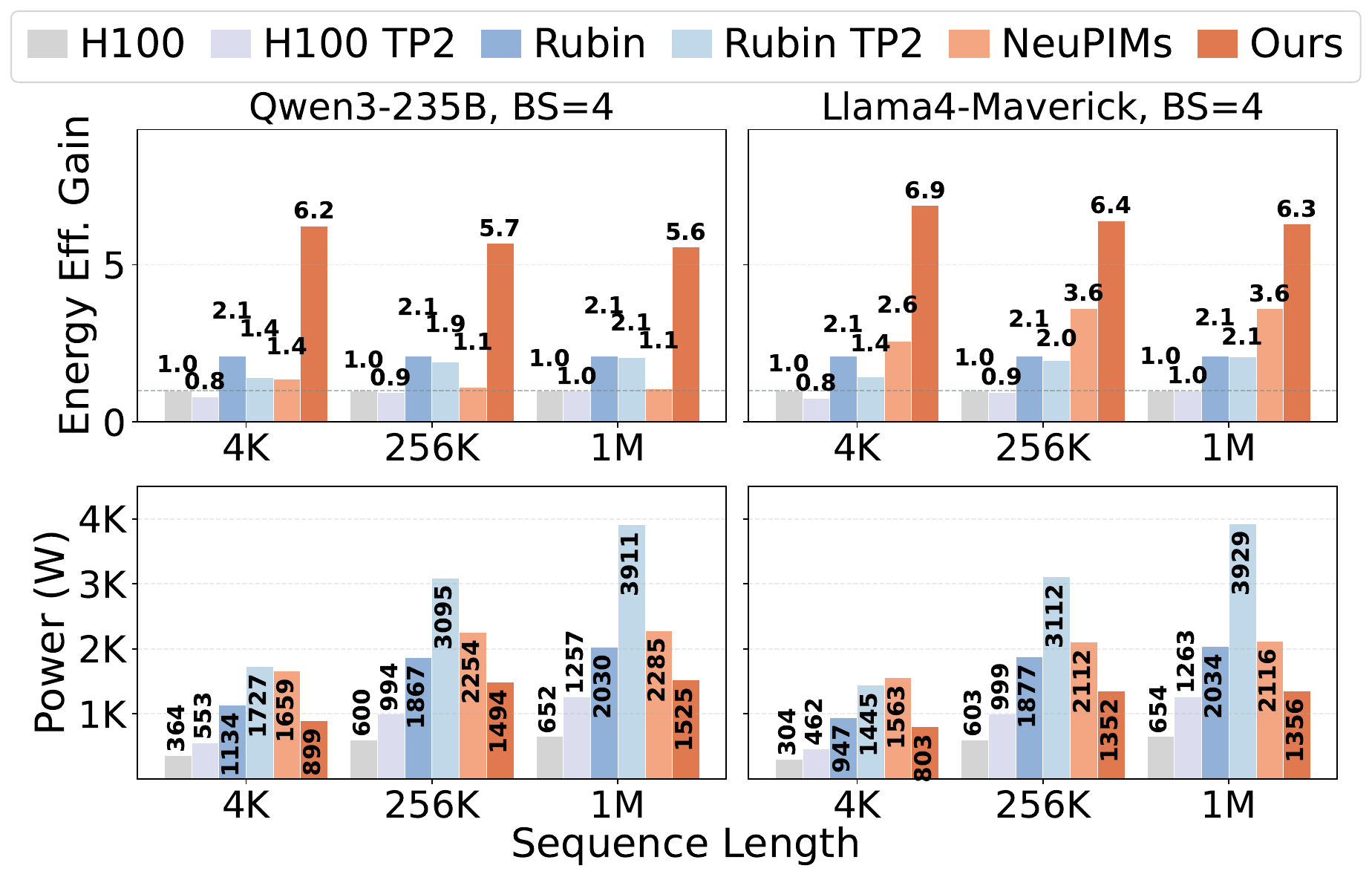}
    \caption{Energy and power analysis.}
    \label{fig:fig2_tokenper}
\end{figure}

%--------------------------------------------------------------------------------%
%--------------------------------------------------------------------------------%

%(multi-column)\\
%Model: Qwen3, Llama4\\
%Seq: 4K, 64K, 256K, 1024K\\
%Compare: Ours(96TFLOPS), Rubin, Rubin TP2, H100, H100 TP2\\
%--------------------------------------------------------------------------------%
\subsection{Latency and System-Level Speedup}
% Keep Original Decode latency Figure (You can compress the figure and delete some data if needed)
% Add a new figure for MLA/CSA/A

% Add a Pareto Frontier fig for MLA. Compare between two of our proposed strategies: Upgrade compute / Add more device.
% Add a figure for different GPU kernel baselines
\color{black}
\textbf{Results on GQA models: }\autoref{fig:fig1_speedup} reports per-layer decode latency speedup across two GQA models, Qwen3 and Llama4, all normalized to a single H100. 
\xname{} achieves the highest speedup everywhere, 12.0--16.3$\times$ over H100 at BS=1 and 13--20$\times$ at BS=32, by packing 16 HBM4 cubes per package versus 5 HBM3e and aggregating 40\,TB/s, 11.9$\times$ that of H100.

Against Rubin (8 HBM4 cubes), we maintain a stable 1.8--2.5$\times$ lead. Even against Rubin TP2, which matches our aggregate bandwidth, we retain 1.5--2.4$\times$ at short-to-medium sequences, narrowing to 1.1$\times$ only at 1M tokens where the large KV-cache volume amortizes Rubin's inter-chip NVLink overhead. Despite this modest gap at 1M, our design delivers 2.8$\times$ better energy efficiency (\autoref{fig:fig2_tokenper}).

Compared with the PIM baseline NeuPIMs, we hold a 3.4$\times$ advantage on Qwen3-235B and 1.4$\times$ on Llama4-Maverick. 
NeuPIMs still relies on a GPU for projection, and its bank-level PIM offers 4.5$\times$ our bandwidth but too little compute, making GQA compute-bound and stranding that bandwidth.

{\textbf{Results on the DeepSeek series.} On DeepSeek-V3 with MLA we observe a markedly different trend.
At short sequences (4K) projection dominates and \xname{} outperforms Rubin by 1.9$\times$ thanks to its bandwidth advantage. As the sequence grows and attention dominates, however, Rubin surpasses \xname{} by up to 2.9$\times$, because MLA has roughly 8$\times$ higher arithmetic intensity than GQA and becomes compute-bound under \xname{}'s per-cube budget.
This is where \xname{}-EC wins: with $4\times$ the compute it regains a 1.8--2.1$\times$ lead over Rubin at a lower power budget.}
% Upgrading the per-cube computing power {from 96 to 384\,TFLOPS by replacing the $16{\times}16$ systolic arrays with $32{\times}32$ arrays} restores \xname{}'s lead to 1.8--2.1$\times$ across all configurations, suggesting that single-chip MLA deployment demands higher compute density. In practice, area and power budget can be reallocated from D2D bandwidth to additional compute units.}

% {
% Deploying MLA across multiple devices lowers latency through parallelism, since DeepSeek V3 has only one KV head and frameworks such as SGLang split the Q heads and replicate the KV cache across devices under TP2--4. Each chip, however, keeps the same compute-to-bandwidth ratio, so a single chip remains compute-bound, and the enhanced-compute Ours-EC is the principled fix that raises per-chip compute directly. Coordinating parallelism across chips is itself a hard problem, because each chip packs 16 cubes and only the edge cubes can link to other chips, which makes the cross-chip topology intricate. We therefore leave multi-chip parallelism to future work and focus here on the single-device strategy centered on the compute upgrade.}

\begin{figure}[t]
    \centering
    \includegraphics[width=0.85\columnwidth]{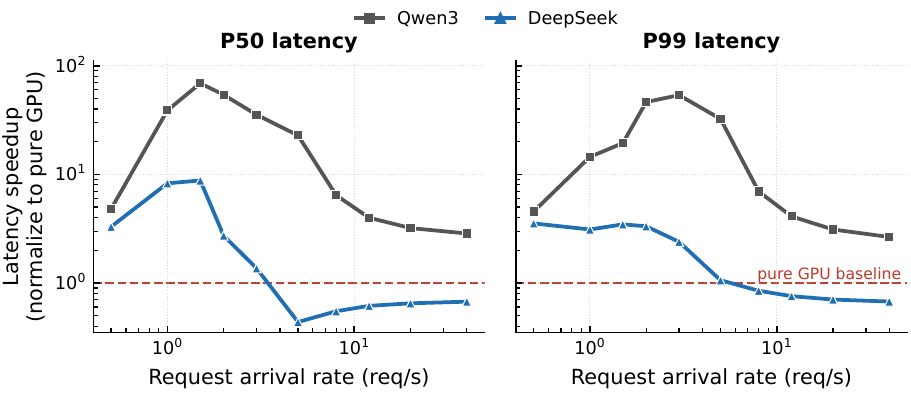}
    \caption{System-level latency for \xname{}-EC.}
    \label{fig:fig1_contbatch}
\end{figure}
%增加Deepseek的结果
{
\textbf{End-to-end decode step.} We evaluate the full decode step in which \xname{} runs the attention and an LPU runs the FFN, broken down in \autoref{fig:fig1_e2e_speedup}.
In the standard setting \xname{} and the LPU are connected by a NIC with 2.4\,$\mu$s latency and 100\,GB/s bandwidth, and in the ideal setting by a Groq-style link with 300\,ns latency and 230\,GB/s bandwidth~\cite{bitarship}.
\xname{}'s attention, projection, and the LPU-side FFN are each far cheaper than on the integrated baselines, giving a 4--5$\times$ speedup over the GPU baseline, although the \xname{}-to-LPU activation transfer becomes the dominant cost, roughly 70\% of the step at 8K but amortizing to about 13\% at 1M.
With a low-overhead interconnect the ideal version reaches a 10--11$\times$ speedup, underscoring the importance of the inter-chip link for low-latency serving.
}
\color{black}

\begin{figure}[t]
    \centering
    \includegraphics[width=\columnwidth]{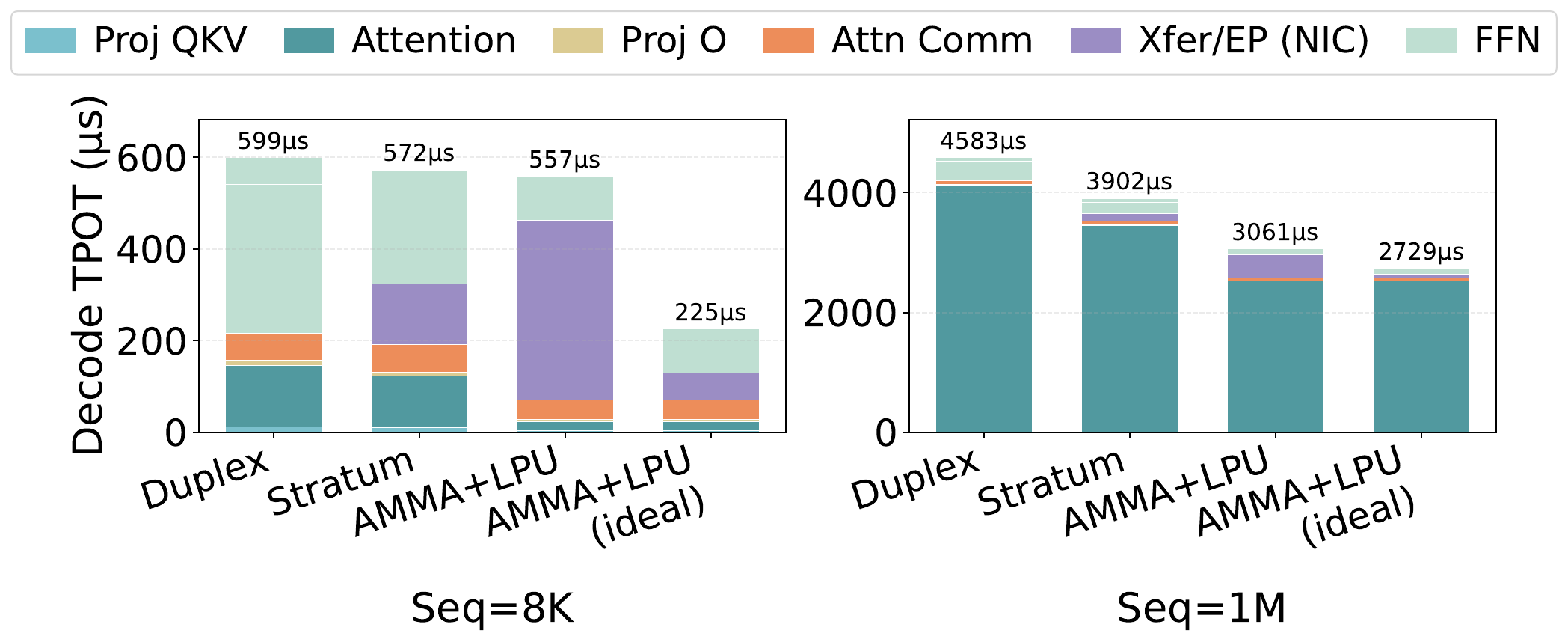}
    \caption{Comparison with 3D-DRAM baselines.}
    \label{fig:fig3_3baseline}
\end{figure}
{
\textbf{Versus 3D-DRAM accelerators.} \autoref{fig:fig3_3baseline} compares \xname{} with the hybrid-bonding accelerator Duplex~\cite{yun2024duplex} and the 3D-DRAM accelerator Stratum~\cite{stratum}. Although both rely on radical future memory technology, \xname{}'s moderate compute and balanced architecture still match them at 8K and run 1.3 to 1.5$\times$ faster at 1M.
Unlike these designs, \xname{} uses mature memory technology with conservative parameters already adopted by the Rubin GPU. Since HBM4 and HBM4E already expose a customizable logic die at an advanced node, \xname{} could be practical within a few years, and its logic-die design philosophy remains compatible with 3D-DRAM should that technology mature.}
\color{black}

\subsection{Energy, Power, and Area Efficiency}

% use a metric: Thougput/package area
\begin{figure}[t]
    \centering
    \includegraphics[width=\columnwidth]{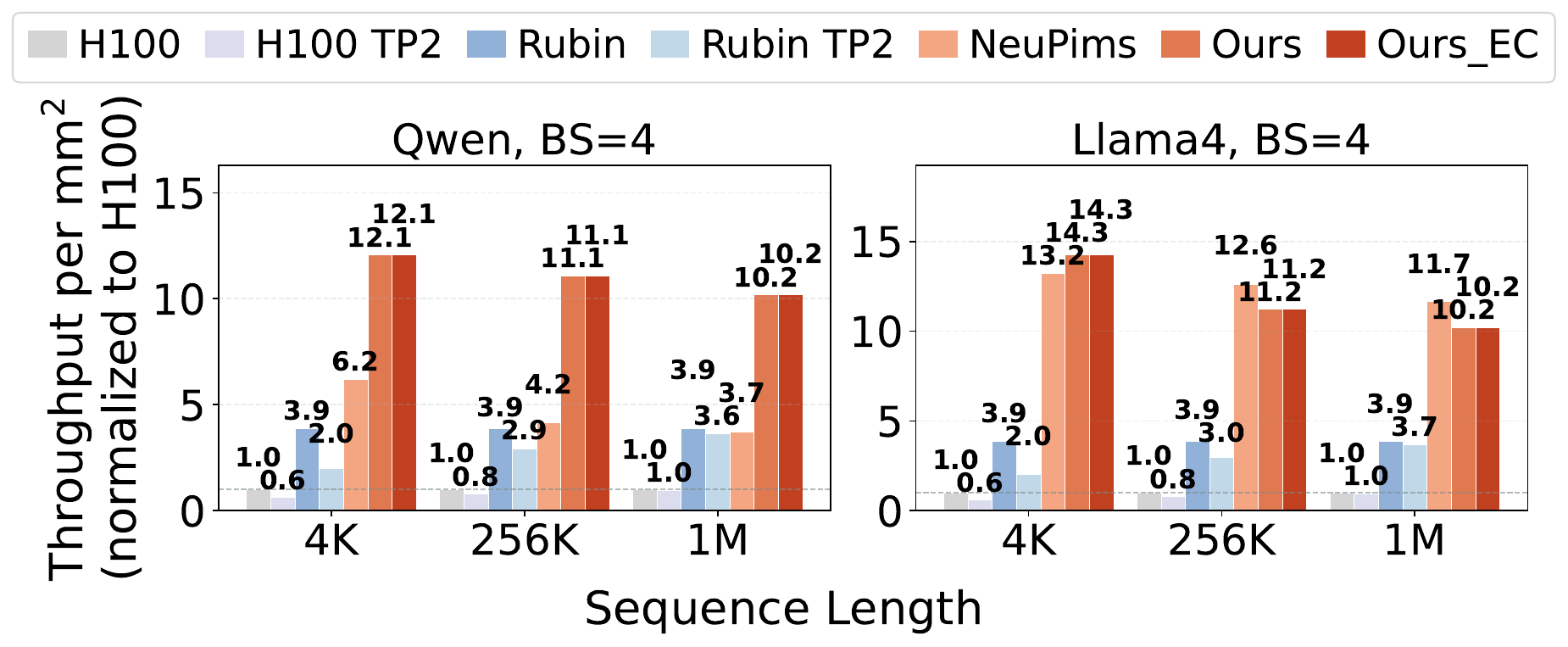}
    \caption{Iso-area throughput.}
    \label{fig:fig2_fig1c}
\end{figure}
% Fig 17 duplicates Fig 16 (iso-area throughput); commented out, may re-add later.
% \begin{figure}[t]
%     \centering
%     \includegraphics[width=\columnwidth]{fig/7_fig1d_iso_area_speedup_bs4_1x2.pdf}
%     \caption{Iso-Area Throughput}
%     \label{fig:fig2_fig1d}
% \end{figure}
{
\textbf{Iso-area throughput.} \autoref{fig:fig2_fig1c} reports throughput per unit package area. 
% Because \xname{} and Rubin occupy a similar package area, this metric isolates architectural efficiency from raw scale. 
\xname{} delivers higher area efficiency, showing that the advantage comes from the architecture rather than a larger package, and reaches about 6 to 7$\times$ the throughput per unit area of Rubin~TP2.
}
\color{black}
% /package size
     
\textbf{Energy and power.} The top of \autoref{fig:fig2_tokenper} shows \xname{} achieves a 2.6--3.1$\times$ energy-efficiency advantage over Rubin, and 2.8$\times$ (at 1M) to 4.8$\times$ (at 4K) over Rubin~TP2, whose doubled static power and NVLink overhead erode efficiency at shorter sequences. This follows because our dedicated microarchitecture provisions only 10\% of Rubin's compute and removes redundant on-chip SRAM. The bottom of \autoref{fig:fig2_tokenper} confirms the power benefit: \xname{} stays below 1500\,W, 1.4$\times$ lower than Rubin and 2.6$\times$ lower than Rubin~TP2, while still outperforming both in absolute terms.

{
\textbf{Area and power breakdown.} \autoref{tab:amma_ec_area_power_single} breaks down the per-cube area and power at a 4\,nm logic-die node, with SRAM modeled in CACTI, the systolic arrays and crossbars in Yosys, and the PHY from open-source data. Standard \xname{} occupies about 52\,mm$^2$ and draws 15\,W, comfortably within the roughly 121\,mm$^2$ logic die. Even the EC variant, with $4\times$ the compute, uses only 92\,mm$^2$ and still fits on the die.}

%--------------------------------------------------------------------------------%
%--------------------------------------------------------------------------------%

\begin{table}[t]

\centering
\scriptsize
\caption{Power and area breakdown (121 mm$^2$ logic die).}
\label{tab:amma_ec_area_power_single}
\resizebox{0.8\columnwidth}{!}{
\begin{tabular}{l|cc|cc}
\hline
\textbf{Component}
& \multicolumn{2}{c|}{\textbf{Area (mm$^2$)}}
& \multicolumn{2}{c}{\textbf{Power (W)}} \\
\cline{2-5}
& \textbf{AMMA} & \textbf{AMMA\_EC}
& \textbf{AMMA} & \textbf{AMMA\_EC} \\
\hline
MAC arrays      & 11.82 & 47.44 & 2.20 & 8.84 \\
SRAM buffers    & 10.01 & 13.01 & 4.05 & 4.05 \\
Two-level xbar  & 2.00  & 3.98  & 0.02 & 0.02 \\
PHY + mem ctrl  & 28.00 & 28.00 & ~9.00 & ~9.00 \\
\hline
\textbf{Total} & \textbf{51.83} & \textbf{92.43} & \textbf{15.27} & \textbf{21.91} \\
\hline
\end{tabular}
}
\end{table}

%--------------------------------------------------------------------------------%
%--------------------------------------------------------------------------------%
\begin{figure}[t]
    \centering
    \includegraphics[width=\columnwidth]{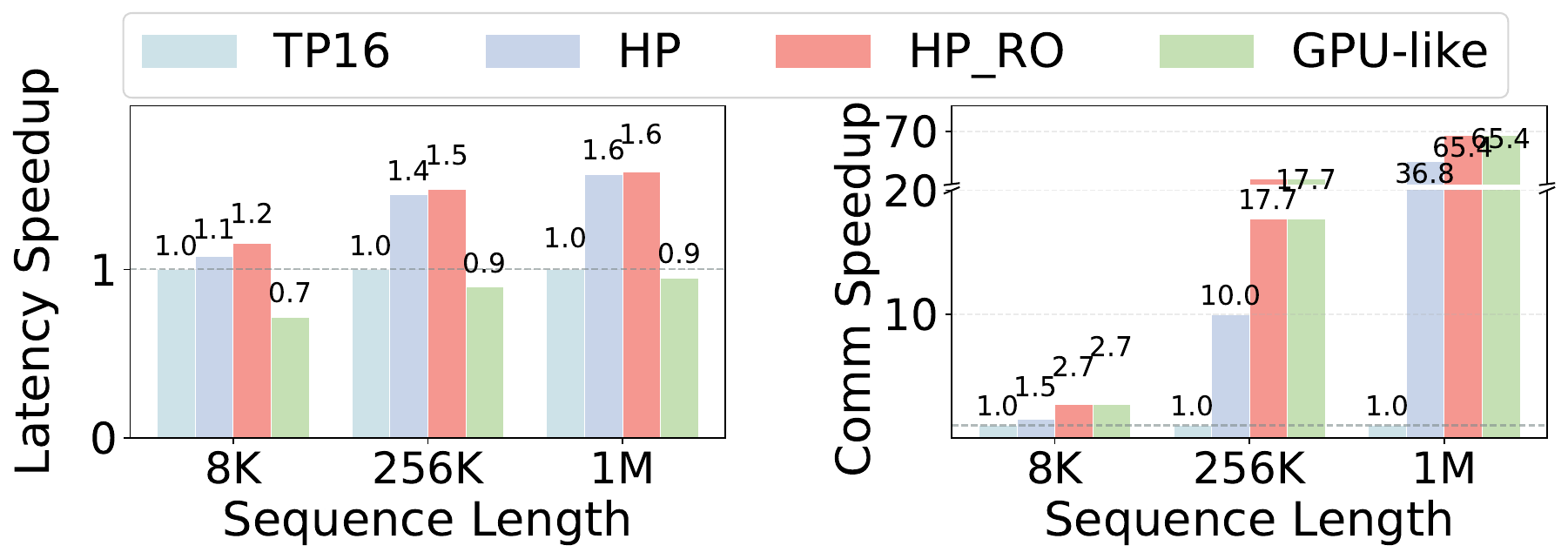}
    \caption{Ablation study.}
    \label{fig:fig3_ablation}
\end{figure}

\subsection{Ablation Study}
% Add a new case, GPU-like arch, with 40\% of computing power.
\color{black}
\begin{figure}[t]
    \centering
    \includegraphics[width=0.8\columnwidth]{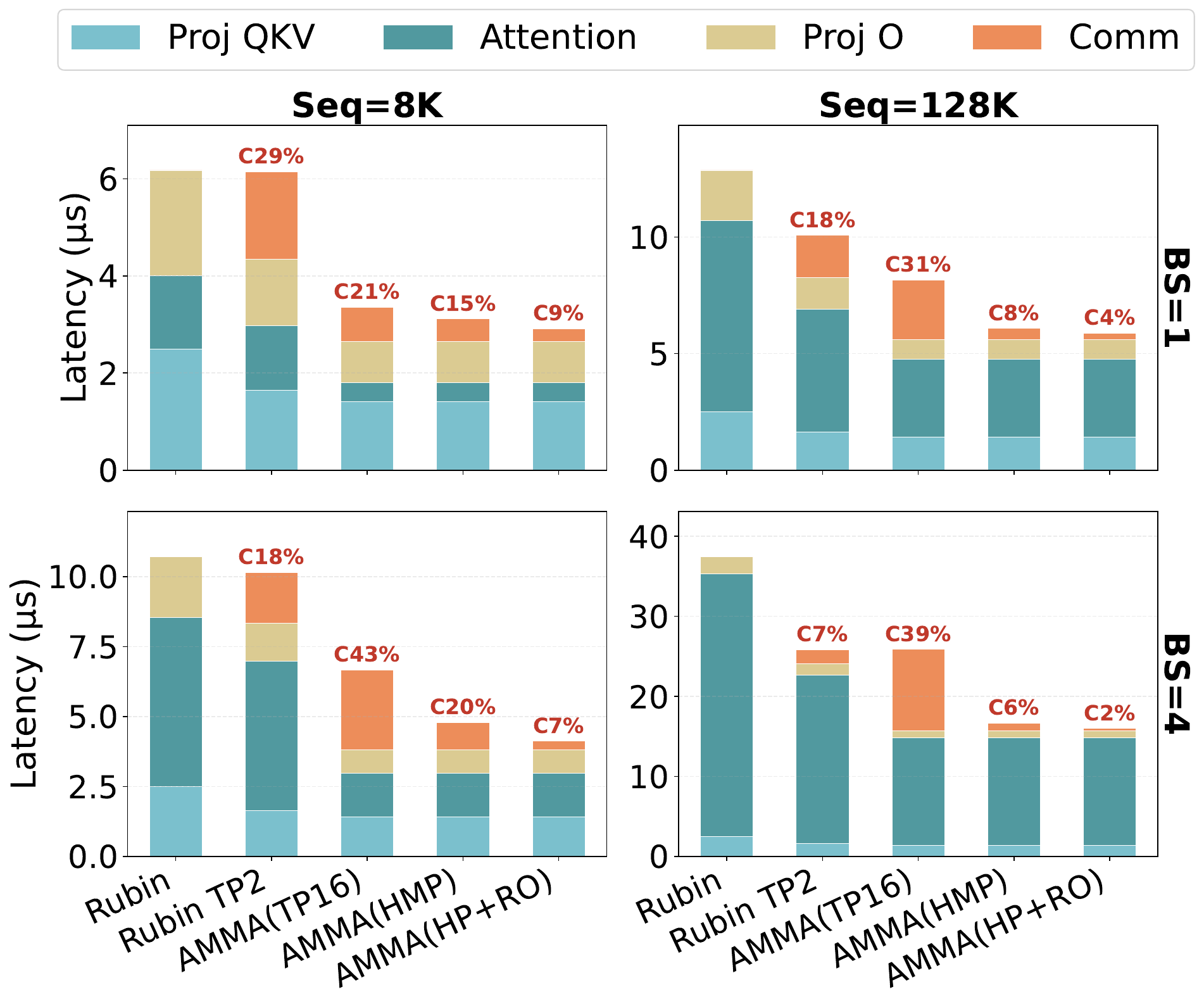}
    \caption{Per-layer decode latency breakdown on Qwen3.}
    \label{fig:fig5_compute_breakdown}
\end{figure}
%add_GPU_like

We run a detailed ablation to isolate the contributions of our parallelism, collective scheduling, and architecture. \textbf{TP16} denotes default GPU-like tensor parallelism across the HBM cubes, \textbf{HP} our two-level hybrid mapping alone, and \textbf{HP\_RO} the full design combining hybrid mapping with reordered collectives. The \textbf{GPU-like} variant runs HP\_RO on a GPU-style multi-level memory hierarchy with limited compute.

As shown in \autoref{fig:fig3_ablation}(a), HP\_RO consistently achieves the highest speedup over the TP16 baseline, reaching 1.1$\times$ at 8K, 1.5$\times$ at 256K, and 1.6$\times$ at 1M. HP also outperforms TP16 but with slightly lower gains (1.1$\times$, 1.4$\times$, and 1.5$\times$). The improvement over TP16 grows with sequence length because TP16 introduces communication volume proportional to the sequence length. Compared with HP, HP\_RO shows a more pronounced advantage at short sequences because RO yields a fixed latency reduction that is increasingly diluted by attention compute as sequences grow.
{
The GPU-like variant reaches only 0.7 to 0.9$\times$ the TP16 baseline, confirming that \xname{}'s gain comes from its memory-centric architecture rather than area or bandwidth alone.}

Since most gains stem from reduced communication, \autoref{fig:fig3_ablation}(b) isolates communication-only speedup. HP\_RO delivers 2.7$\times$, 17.7$\times$, and 65.4$\times$ communication speedup at 8K, 256K, and 1M, respectively, compared to 1.3$\times$, 10.0$\times$, and 36.8$\times$ for HP alone.
%--------------------------------------------------------------------------------%
%--------------------------------------------------------------------------------%

\begin{figure}[t]
    \centering
    \includegraphics[width=\columnwidth]{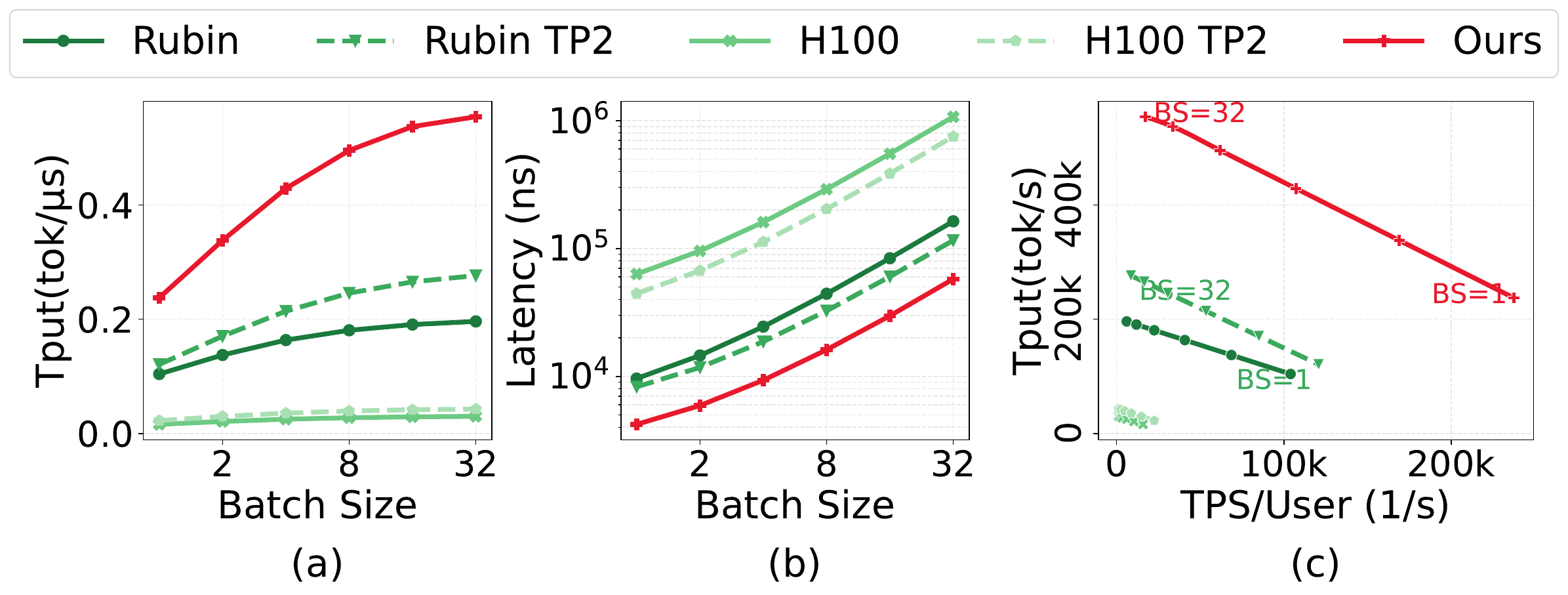}
    \caption{Batch-size exploration for Qwen3-235B at seq=64K: (a) throughput and (b) end-to-end latency vs. batch size, (c) throughput vs. per-user speed (TPS/User).}
    \label{fig:fig4_combined}
\end{figure}

\subsection{Batch Size and Time Breakdown}
\textbf{Batch size.} Since low-latency serving typically involves batch sizes of 1--32, we study how batch size affects the latency-throughput trade-off.
As shown in~\autoref{fig:fig4_combined}(a) and (b), when the batch size increases from 1 to 32, total system throughput improves from 0.223 to 0.478\,tok/$\mu$s (2.14$\times$), but latency also worsens by 30$\times$.
The throughput improvement arises because batching increases the arithmetic intensity of projection operations, allowing~\xname{}'s computing power to be more fully utilized.
The 30$\times$ latency rise occurs because attention does not benefit from batching: each query attends only to its own KV cache, so requests run sequentially and output projection waits for the whole batch.

The flattening trend of the red line in~\autoref{fig:fig4_combined}(a) further indicates that the computing power of~\xname{} is already saturated when the batch size reaches 16. Increasing the batch size beyond this point yields no further throughput improvement.
This contrasts with GPUs, which need a batch of 200--400 to saturate compute.

The Pareto frontier in~\autoref{fig:fig4_combined}(c) compares~\xname{} with NVIDIA GPUs.
Across batch sizes,~\xname{} consistently beats GPUs on long-context attention, with 14--16$\times$ higher throughput than H100 at comparable or lower power.

%--------------------------------------------------------------------------------%
%--------------------------------------------------------------------------------%
\textbf{Per-stage time breakdown.} \autoref{fig:fig5_compute_breakdown} breaks per-layer decode latency into Proj QKV, Attention, Proj O, and Communication.

At short sequences, projection dominates compute time: at BS=1 and Seq=8K, Proj QKV and Proj O together account for 85\% of AMMA(HP+RO)'s compute, while attention contributes only 15\%. As sequence length grows to 128K, attention scales proportionally and becomes dominant, reaching 60\% at BS=1 and 86\% at BS=4, since each batch item's KV cache must be streamed independently. By distributing weights and KV cache across 16 HBM cubes with PNM, AMMA achieves 40\,TB/s aggregate bandwidth (vs.\ Rubin's 22\,TB/s), reducing Proj QKV by 1.76$\times$, Proj O by 2.57$\times$, and attention by 3.89$\times$ compared to Rubin.

\subsection{Design-Space Exploration}

\begin{figure}[t]
    \centering
    \includegraphics[width=0.8\columnwidth]{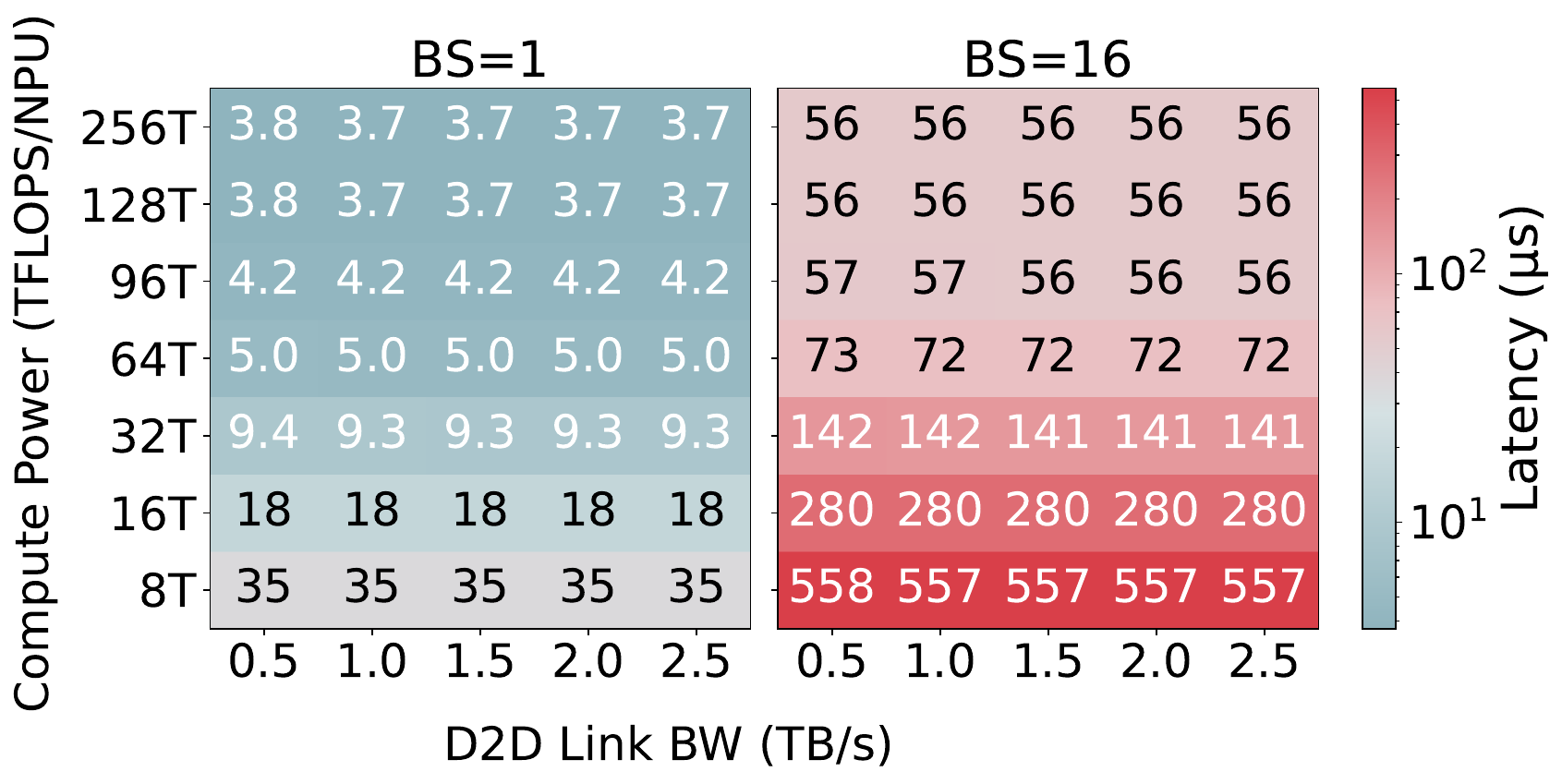}
    \caption{Hardware parameter exploration on Qwen3-235B.}
    \label{fig:fig6_heatmap}
\end{figure}

{
% Consider simplifying existing figures, for example, only keep BS=1/16
% Add sensitivity study figure on D2D Latency. 
%增加折线图D2D latency 
\textbf{Compute power vs.\ D2D bandwidth.} We explore how per-cube compute power and D2D bandwidth affect total latency, offering hardware designers guidance for navigating area and power trade-offs on the logic die. We sweep compute power from 8 to 256\,TFLOPS and D2D link bandwidth from 0.5 to 2.5\,TB/s. Results are shown in \autoref{fig:fig6_heatmap}, where blue indicates low latency and red indicates high.
Compute power matters more than D2D bandwidth for two reasons. 
First, our hybrid parallelism and reordered collectives already minimize inter-cube communication. 
Second, at low batch sizes (1--32) the D2D volume is small and latency is dominated by fixed startup delay, so higher bandwidth helps little.}

\begin{figure}[t]
    \centering
    \includegraphics[width=0.8\columnwidth]{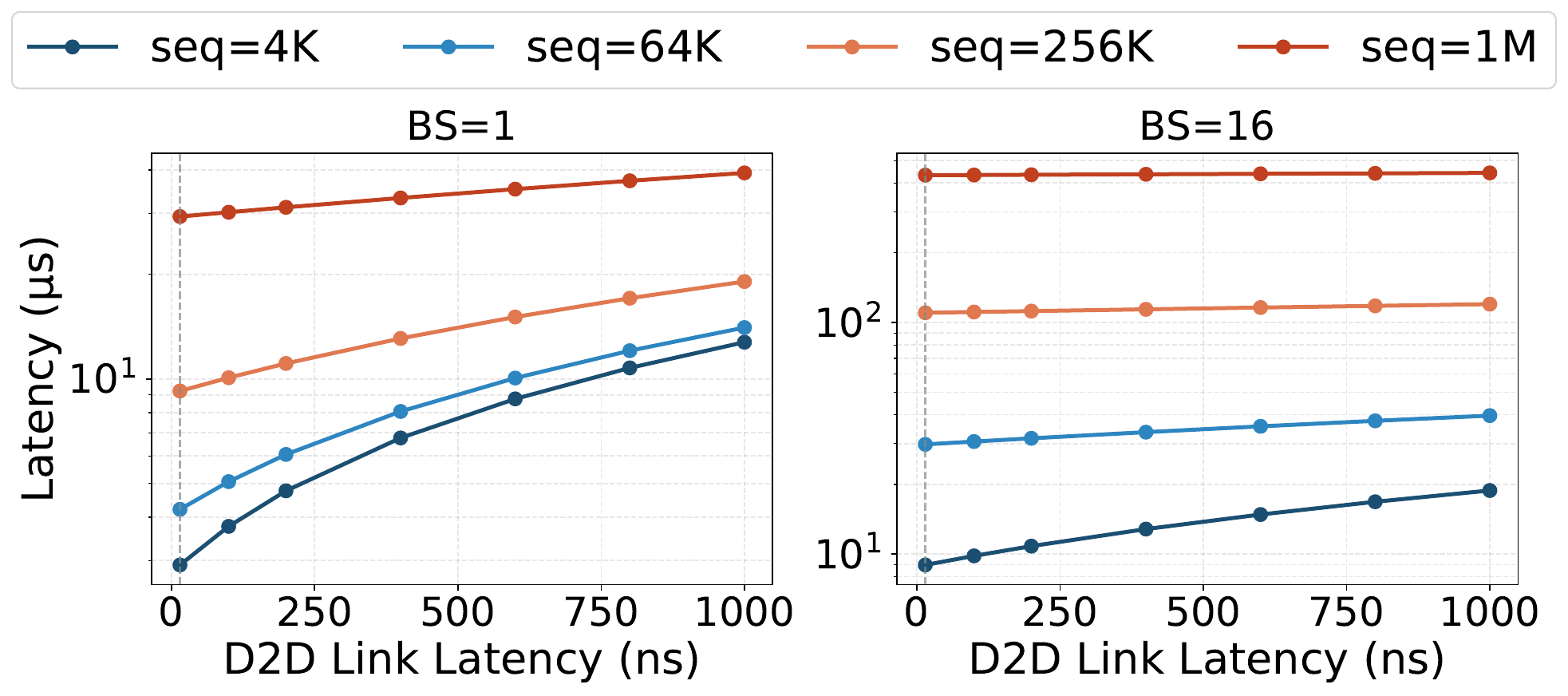}
    \caption{D2D latency sensitivity.}
    \label{fig:fig6_d2d_latency}
\end{figure}

{
\textbf{D2D link latency.} \autoref{fig:fig6_d2d_latency} sweeps the D2D link latency. End-to-end latency is highly sensitive to it at small batch, underscoring the importance of a low-latency D2D protocol for extremely low-latency, small-batch serving.}

\begin{figure}[t]
    \centering
    \includegraphics[width=0.95\columnwidth]{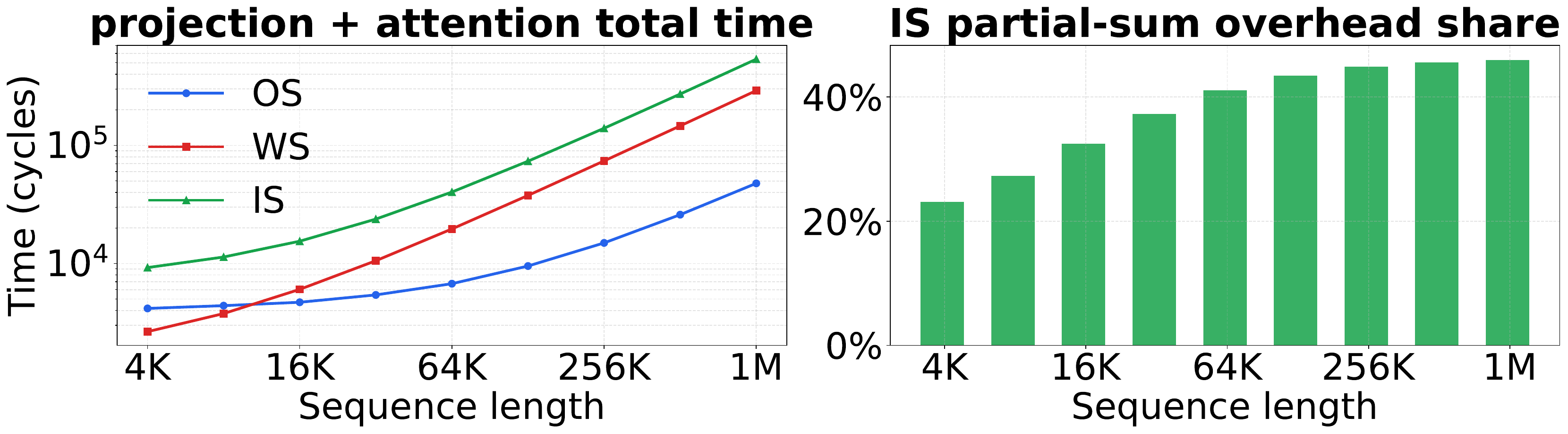}
    \caption{Dataflow exploration on Qwen3-235B.}
    \label{fig:fig7_dataflow}
\end{figure}

{
\textbf{Systolic-array dataflow.} We use a single fixed dataflow, since heterogeneous dataflows add routing, multiplexing, buffering, and control logic that cost energy and area~\cite{han2024redas, kwon2018maeri, kwon2021heterogeneous}. Among output-stationary (OS), input-stationary (IS), and weight-stationary (WS) dataflows (\autoref{fig:fig7_dataflow}), OS achieves the lowest latency at long sequences, since its partial sums stay local until accumulation completes and avoid the psum reduction that IS and WS incur.}
\color{black}

% Orphan figure (never \autoref'd); commented out to save space. Re-add + cite for D2D-latency sensitivity if wanted.

\section{Related Work}

\noindent\textbf{GPU-centric long-context serving:}
Many works improve long-context serving on GPUs~\cite{dai2025flashdecoding++next,hong2024flashdecoding++,pagedattention,zhong2024distserve,lin2024infinite,wu2024loongserve,zhu2025megascale,wang2025step}: the FlashDecoding series optimizes decode-attention kernels~\cite{hong2024flashdecoding++,dai2025flashdecoding++next}, DistServe, Infinite-LLM, and LoongServe scale it across GPU clusters~\cite{zhong2024distserve,lin2024infinite,wu2024loongserve}, and MegaScale-Infer and Step-3 disaggregate attention from FFN onto separate pools~\cite{zhu2025megascale,wang2025step}. These stay GPU-centric and keep the mismatch between compute-rich GPUs and memory-bound decode attention.

\noindent\textbf{PIM/NMP architecture:}
Many works move computation into or near memory~\cite{upmem,LP-Spec,kal2021space,ke2020recnmp,kim2025nmp-pak,kwon2019tensordimm,aim-gddr6,lee2021hbm-pim,yun2022grande}: PIM designs across GDDR, DDR, HBM, and LPDDR accelerate bandwidth-limited kernels such as embedding lookups and reductions~\cite{upmem,aim-gddr6,lee2021hbm-pim,kwon2019tensordimm,kal2021space,LP-Spec}, while NMP designs such as RecNMP~\cite{ke2020recnmp}, NMP-PaK~\cite{kim2025nmp-pak}, and GraNDe~\cite{yun2022grande} target recommendation, graph, and genome workloads.

\noindent\textbf{Memory-side architectures for LLM:}
Memory-side PIM and PNM also apply to LLM attention. One line pushes attention near memory with PIM, including AttAcc, TransPIM, LP-Spec, AttenPIM, and the long-context LoL-PIM~\cite{attacc,zhou2022transpim,LP-Spec,chen2025attenpim,kwon2024lol-pim}. Another reduces attention cost by memory-side filtering or retrieval, as in DReX~\cite{DReX} and LongSight~\cite{quinn2025longsight}, which fetch only the most relevant distant keys while leaving dense attention on the GPU. Both still rely on the GPU as a central hub for communication and scheduling.
% {A recent line co-designs dedicated memory-centric serving accelerators on emerging substrates, including the wafer-scale WaferLLM~\cite{waferllm}, the 3D-stacked-DRAM Stratum~\cite{stratum} and HeLioS~\cite{helios}, and the HBM-PIM Duplex~\cite{yun2024duplex}. Several of these assume memory technologies with no existing prototype, whereas \xname{} runs on the customizable logic die that HBM4 already exposes.}
In contrast, \xname{} replaces GPU compute dies with HBM-NMP cubes and treats the memory-side device as a first-class accelerator, opening a new avenue for memory-centric architecture.

\section{Conclusion}

We presented \xname{}, a multi-chiplet HBM-PNM architecture that replaces GPU compute dies with PNM-enabled HBM cubes to serve as a standalone, memory-centric accelerator for long-context decode attention.
Through co-design of a dedicated logic-die microarchitecture, two-level hybrid parallelism, and reordered collectives, \xname{} fully exploits aggregated HBM bandwidth while keeping inter-cube data movement minimal, and our exploration reveals actionable compute and bandwidth trade-offs under area and power limits.
We hope this work spurs the community to explore memory-centric architectures for future disaggregated serving systems.

% We presented \xname{}, a multi-chiplet HBM-PNM architecture that replaces GPU compute dies with PNM-enabled HBM cubes to serve as a standalone, memory-centric accelerator for long-context decode attention. By co-designing a dedicated logic-die microarchitecture, a two-level hybrid parallelism scheme, and a reordered collective communication flow, \xname{} fully exploits the aggregated internal HBM bandwidth while keeping inter-cube data movement local and minimal. Our design-space exploration further reveals actionable trade-offs between per-cube compute throughput and inter-die link bandwidth under realistic area and power constraints. We hope this work demonstrates that memory-centric processors can stand as a distinct, first-class device class alongside GPUs and LPUs, and encourages the community to explore heterogeneous serving architectures beyond the GPU-centric paradigm.

\section*{Acknowledgment}
We thank all reviewers for their constructive feedback and insightful suggestions. This work is partially supported by Samsung Semiconductor.

\appendix[Artifact Appendix]

%%%%%%%%%%%%%%%%%%%%%%%%%%%%%%%%%%%%%%%%%%%%%%%%%%%%%%%%%%%%%%%%%%%%%
\subsection{Abstract}

This artifact reproduces the paper's evaluation-section figures for \emph{AMMA}, an HBM near-memory-processing attention accelerator for LLM decode. It packages about 16{,}500 lines of pure-Python analytical modeling, a roofline built on profiled utilization together with a closed-form power model, plus the checked-in intermediate data and the plotting scripts. The figures compare AMMA's decode latency, energy, and power against NVIDIA H100 and Rubin GPUs and a published near-memory accelerator across GQA, MLA, HCA, and CSA attention at contexts up to 1M tokens. To keep evaluation lightweight, the two cycle-level simulators used in the paper, ScaleSim and AstraSim, are replaced by fast analytical backends, and live GPU profiling is replaced by pre-captured measurements. A reviewer therefore needs no simulator and no GPU, and a single command \texttt{python3 quickstart.py} runs the tests and regenerates every figure from checked-in data.

\subsection{Artifact check-list (meta-information)}

{\small
\begin{itemize}
  \item {\bf Algorithm: } analytical roofline modeling of decode attention, composed with a profiled-utilization model and a closed-form power model.
  \item {\bf Program: } the AMMA decode-modeling framework, about 16{,}500 lines of pure Python~3.
  \item {\bf Model: } Qwen3-235B, DeepSeek-V3, and Llama-4 decode workloads, modeled analytically from published configurations. No model weights are needed.
  \item {\bf Data set: } a trace-derived distinct-expert activation curve from open-source Qwen3-235B and DeepSeek-R1 MoE traces, plus H100 and Rubin profiling snapshots. All bundled.
  \item {\bf Run-time environment: } Linux with Python~3, tested on 3.10 and 3.13. NumPy, SciPy, Matplotlib, and pytest, installed automatically.
  \item {\bf Hardware: } any CPU host with under 1\,GB of RAM. No GPU.
  \item {\bf Execution: } a single command, \texttt{python3 quickstart.py}. Single-threaded.
  \item {\bf Metrics: } decode-attention latency, energy efficiency in Token/J, power, and throughput, each normalized to the H100 baseline.
  \item {\bf Output: } nine figures in PDF, named after the paper figures, plus CSV and JSON intermediate results.
  \item {\bf Experiments: } the nine evaluation-section figures, Figures~11, 12, 13, 14, 15, 17, 18, 19, and 20, regenerated by a single command.
  \item {\bf How much disk space required (approximately)?: } about 20\,MB.
  \item {\bf How much time is needed to prepare workflow (approximately)?: } a few minutes for the one-time dependency install.
  \item {\bf How much time is needed to complete experiments (approximately)?: } one to two minutes, about 15 seconds on a warm machine.
  \item {\bf Publicly available?: } yes, on GitHub.
  \item {\bf Code licenses (if publicly available)?: } Apache-2.0.
  \item {\bf Data licenses (if publicly available)?: } Apache-2.0.
  \item {\bf Workflow automation framework used?: } a self-contained Python driver, \texttt{quickstart.py}.
  \item {\bf Archived (provide DOI)?: } \url{https://doi.org/10.5281/zenodo.21540779}
\end{itemize}
}

%%%%%%%%%%%%%%%%%%%%%%%%%%%%%%%%%%%%%%%%%%%%%%%%%%%%%%%%%%%%%%%%%%%%%
\subsection{Description}

\subsubsection{How to access}

The artifact is archived on Zenodo at \url{https://doi.org/10.5281/zenodo.21540779} and developed on GitHub at \url{https://github.com/zhongkaiyu/AMMA}. The \texttt{README.md} documents setup and data flow, and the Zenodo record snapshots the evaluated commit.

\subsubsection{Hardware dependencies}

A standard CPU host is sufficient and no GPU is required. The default workflow uses under 1\,GB of RAM, with a peak near 200\,MB, and runs on a single core because the pipeline is single-threaded.

\subsubsection{Software dependencies}

Python~3 with NumPy, SciPy, Matplotlib, and pytest, whose minimum versions are pinned in \texttt{requirements.txt} and installed automatically by the quickstart script. No proprietary software is required. Optional regeneration of the vendored inputs additionally needs the ScaleSim and AstraSim (with Chakra) simulators.

\subsubsection{Data sets}

All data required for evaluation is bundled. The end-to-end and continuous-batching experiments use a trace-derived distinct-expert activation curve summarized from open-source Qwen3-235B and DeepSeek-R1 MoE expert-selection traces. The attention experiments use checked-in H100 and Rubin profiling snapshots. Nothing external needs to be downloaded.

\subsubsection{Models}

The decode workloads for Qwen3-235B, DeepSeek-V3, and Llama-4 are modeled analytically from their published configurations, so no model weights are needed.

%%%%%%%%%%%%%%%%%%%%%%%%%%%%%%%%%%%%%%%%%%%%%%%%%%%%%%%%%%%%%%%%%%%%%
\subsection{Installation}

Clone the repository or download the Zenodo archive. No build step and no manual dependency install are required, because \texttt{quickstart.py} bootstraps its own environment. On a host that lacks the Python packages it creates a local virtual environment, installs \texttt{requirements.txt}, and continues inside it automatically.

%%%%%%%%%%%%%%%%%%%%%%%%%%%%%%%%%%%%%%%%%%%%%%%%%%%%%%%%%%%%%%%%%%%%%
\subsection{Experiment workflow}

The command \texttt{python3 quickstart.py} runs the unit-test gate and regenerates the nine evaluation-section figures into \nolinkurl{figures_reproduced/}, each written under the name it carries in the paper such as \nolinkurl{figure11_decode_speedup.pdf}. A single figure renders with \texttt{-f <key>}, where the key is the paper figure number, and \texttt{-{}-list} prints every key with its output file. The gate also validates the two analytical replacements against their simulators, since \nolinkurl{tests/test_comm_model_vs_astrasim.py} confirms the communication model matches the AstraSim data for the AMMA strategies, and \nolinkurl{scalesim/test_scalesim.py} checks the systolic-array utilization against ScaleSim.

%%%%%%%%%%%%%%%%%%%%%%%%%%%%%%%%%%%%%%%%%%%%%%%%%%%%%%%%%%%%%%%%%%%%%
\subsection{Evaluation and expected results}

Success means the console reports \texttt{9 passed} and \texttt{reproduced: 9/9}. Each output in \nolinkurl{figures_reproduced/} is named after the paper figure it reproduces, so a reviewer matches them directly. The artifact regenerates Figures~11, 12, 13, 14, 15, 17, 18, 19, and 20. The model is deterministic, so the figures reproduce with no run-to-run variation. Consistent with the paper, AMMA achieves up to about 15.5$\times$ lower decode-attention latency and about 6.9$\times$ lower energy than the H100.

%%%%%%%%%%%%%%%%%%%%%%%%%%%%%%%%%%%%%%%%%%%%%%%%%%%%%%%%%%%%%%%%%%%%%
\subsection{Experiment customization}

Individual figures can be rendered with \texttt{-f}, and the analytical backends under \nolinkurl{backend/} accept different model and hardware presets on the command line. The merge step exposes a \texttt{-{}-comm-model} flag that selects the analytical roofline communication model, which is the default and needs no simulator, or communication vendored from AstraSim.

%%%%%%%%%%%%%%%%%%%%%%%%%%%%%%%%%%%%%%%%%%%%%%%%%%%%%%%%%%%%%%%%%%%%%
\subsection{Notes}

The vendored AstraSim data and the ScaleSim utilization profiles carry checked-in comparison scripts, so the accuracy of the analytical replacements can be audited without installing either simulator.

%%%%%%%%%%%%%%%%%%%%%%%%%%%%%%%%%%%%%%%%%%%%%%%%%%%%%%%%%%%%%%%%%%%%%
\subsection{Methodology}

Submission, reviewing and badging methodology:

\begin{itemize}
  \item \url{https://www.acm.org/publications/policies/artifact-review-and-badging-current}
  \item \url{https://cTuning.org/ae}
\end{itemize}

\bibliographystyle{IEEEtran}
\bibliography{main}

\end{document}